\documentclass{article}

\usepackage{arxiv}

\usepackage[utf8]{inputenc} 
\usepackage[T1]{fontenc}    
\usepackage{hyperref}       
\usepackage{url}            
\usepackage{booktabs}       
\usepackage{amsfonts}       
\usepackage{nicefrac}       
\usepackage{microtype}      
\usepackage{lipsum}
\usepackage{fancyhdr}       
\usepackage{graphicx}       
\graphicspath{{media/}}     

\renewcommand{\shorttitle}{}

\usepackage{amsmath, amssymb, amsfonts, amsthm}
\usepackage{bm}
\usepackage[normalem]{ulem} 
\usepackage{caption}
\usepackage{subcaption}
\usepackage{graphicx}
\usepackage{tikz}
\usetikzlibrary{arrows, shapes.misc, positioning, decorations.pathmorphing}
\usepackage[authoryear]{natbib}
\usepackage{comment}
\usepackage{arydshln} 

\usepackage{rotfloat}

\floatstyle{ruled}
\newfloat{algorithm}{tbp}{loa}
\providecommand{\algorithmname}{Algorithm}
\floatname{algorithm}{\protect\algorithmname}

\newcommand{\tr}{\operatorname{tr}}
\newcommand{\neighbours}{\operatorname{ne}}

\newcommand{\laplacian}{\operatorname{\operatorname{div}\nabla}}
\newcommand{\KLD}{D_\text{KL}}

\newcommand{\covsample}{\bm{\Sigma}^*}
\newcommand{\E}[1]{\mathbb{E}\left[#1\right]}

\newtheorem{example}{Example}

\definecolor{notused}{rgb}{0,0,1}
\colorlet{notused}{notused!40} 
\definecolor{used}{rgb}{1,0,0}
\colorlet{used}{used!90} 
\newcommand{\argused}[1]{\textcolor{used} {#1}}
\newcommand{\argnotused}[1]{\textcolor{notused} {#1}}

\usepackage{tcolorbox}
\newtcolorbox{highlight}[1]{%
  colback=gray!5!white,
  colframe=gray!30!black,
  fonttitle=\bfseries,
  title=#1
}

\definecolor{ForestGreen}{RGB}{1,68,33}
\definecolor{DarkGreen}{RGB}{0,100,0}
\definecolor{mygreen}{RGB}{1,100,33}
\definecolor{mattgreen}{RGB}{10,140,10}
\definecolor{mattred}{RGB}{255,10,10}
\definecolor{mattblue}{RGB}{10,20,255}

\title{An Ensemble Information Filter:\\
\Large{Learning and Encoding Sparse Structures in Ensemble Kalman Methods}\\
}

\author{
	\href{https://orcid.org/0000-0002-6220-5680}{\includegraphics[scale=0.06]{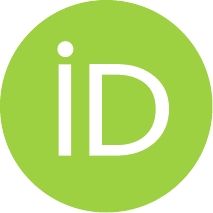}\hspace{1mm}Berent Å.~S.~Lunde}$^{1,2}$\\[2mm]
	$^1$Department of Informatics\\
	University of Bergen\\
	Bergen, Norway \\
    $^2$Equinor, Bergen, Norway\\
	\texttt{berent.a.lunde@uib.no}\\
}

\begin{document}
\maketitle

\begin{abstract}
Ensemble-based Data Assimilation faces significant challenges in high-dimensional systems due to spurious correlations and ensemble collapse.
We propose the Ensemble Information Filter, which addresses these issues by encoding locality directly in the precision matrix through sparse Markov structure. This approach is motivated by connections to SPDE-based models, where local operators induce sparse precision representations and provide a principled form of regularisation.
The framework further incorporates sparse regression for learning the observation operator, enabling stable inference in regimes where $p \gg n$. By enforcing structure prior to estimation, EnIF avoids the need for heuristic localisation and improves both statistical and computational scalability.
The methodology is demonstrated on filtering, smoothing, and parameter estimation problems, where it adapts to the underlying dependence structure and yields stable and efficient inference in high-dimensional systems.
\end{abstract}



\section{Introduction}
\label{sec:introduction}

Ensemble-based Data Assimilation (EDA) is a cornerstone methodology for integrating observations into dynamical models, particularly in high-dimensional non-linear systems. 
It sees applications to weather forecasting and meteorology \citep{houtekamer2005ensemble}, oceanography \citep{evensen1994sequential}, and reservoir engineering \citep{aanonsen2009ensemble}. Despite its widespread success, EDA faces critical statistical challenges such as spurious correlations and ensemble collapse. These are typically addressed through localisation techniques  \citep{anderson2003local,evensen2003ensemble,ott2004local, hamill2001distance, houtekamer2001sequential}. The techniques typically require manual tuning and were introduced as ad-hoc remedies rather than arising from an explicit statistical model.

This paper makes several contributions toward statistically efficient EDA algorithms that are adaptive, computationally scalable, and easy to use for practitioners.

The first contribution is a statistical learning perspective on EDA. 
Following \citet{marzouk2016introduction}, EDA can be viewed as approximating a target distribution $P$ with a model $Q$ by minimising the Kullback-Leibler divergence (KLD). 
When the model $Q$ is estimated from a finite ensemble, this becomes a standard statistical estimation problem \citep{hastie2009elements}. 
From this perspective, classical methods such as the Ensemble Kalman Filter (EnKF) \citep{evensen1994sequential, burgers1998analysis} and Ensemble Smoother (ES) \citep{evensen2009data} implicitly estimate dense dependence structures with a number of parameters that grows quadratically in dimension. 
Specifically, the optimism of the training loss—arising from estimating a high-dimensional dependence structure from a finite ensemble—corresponds to spurious correlations, ensemble collapse, and eventual divergence from $P$.
Addressing these issues requires constraining the statistical model $Q$ a-priori or adaptively during estimation, balancing model complexity with statistical reliability.


Secondly, we encode conditional independence structure directly in the statistical model $Q$. 
This includes classical Markov properties in the state, as well as sparsity in the regression mapping from state to observations when the observation relationship must be learned.
Conditional independence in the state is well studied in Gaussian Markov random fields (GMRF) \citep{rue2005gaussian}, is linked to SPDE dynamics \citep{rue2009approximate}, is recognized in triangular transport \citep{spantini2018inference} and triangular transport with Gaussian processes \citep{katzfuss2024scalable}. Typically this conditional independence is expressed through sparse Cholesky factorisations \citep{rue2005gaussian, schafer2021sparse}.
To the best of our knowledge, the study of learning sparsity also in the observation operator has however received little or no attention.
The ability to learn such estimators is important for practical applicability, as relationships between parameters, states, and observations are often complex or not explicitly specified. 
Imposing sparsity in this mapping also provides interpretability, enabling both global insight into parameter–observation relationships and decomposition of individual updates, which is of practical importance to practitioners.
Encoding these properties into $Q$ via sparsity constraints mitigates spurious correlations while enhancing scalability by reducing computational complexity \citep{rue2009approximate, kristensen2015tmb}, and opens up for explainability.

Thirdly, the paper presents the Ensemble Information Filter (EnIF), a novel approach to EDA that directly encodes conditional independence into the statistical model through sparsity in precision matrices and sparse linear maps. 
Analogous to how the EnKF is an ensemble extension of the Kalman filter \citep{kalman1960new}, the EnIF can be viewed as an ensemble extension of the less known but mathematically equivalent Information filter \citep{kalman1961new, moore1979optimal}.
The reparametrization works internally, and makes it natural to embed the aforementioned modern statistical approaches of spatio-temporal models and high-dimensional regression.
This distinguishes the EnIF from the EnKF, Ensemble Smoother (ES) \citep{evensen2009data}, and related classical EDA approaches, in particular for the finite ensemble setting.
The author considers it a direct competitor to above methods, offering advantages in statistical convergence, adaptiveness, scalability and interpretability.

The paper is laid out as follows: Section \ref{sec:ensemble-based-data-assimilation} introduces EDA and necessary background theory and Section \ref{sec:markov-structure-from-local-operators} discusses Markov properties for filtering and smoothing under SPDE dynamics. Section \ref{sec:an-ensemble-information-filter-encoding-Markov} discusses locality parametrisations and introduces the EnIF EDA algorithm. Section \ref{sec:synthetic-numerical-experiments} provides experiments focusing on various properties of the methodology, while Section \ref{sec:applications} showcase the applicability of the methodology to smoothing, filtering, and parameter estimation. Section \ref{sec:discussion} provides discussion.

\section{Ensemble based data assimilation}\label{sec:ensemble-based-data-assimilation}

\begin{table}[ht]
\centering
\caption{Nomenclature separated in mathematical, problem setup, statistical, and ensemble-representation blocks}
\begin{tabular}{rl}
\hline
\textbf{Symbol} & \textbf{Description} \\
\hline
$x$ & scalar \\
$\bm{x}$ & vector (bold font) \\
$\bm{X}$ & matrix (capitalised bold font) \\

$\tr(\cdot)$ & the trace operator $\tr\left(\bm{X}\right) = \sum_j \bm{X}_{j,j}$ \\
$\Delta, \nabla, \nabla^2$ & finite differences, gradient, and Hessian operators \\
$\laplacian$ & the Laplacian operator as divergence of the gradient field \\

$\mathcal{G} = (\mathcal{V}, \mathcal{E})$ & graph with vertices $\mathcal{V}$ and edges $\mathcal{E}$ \\
$\neighbours(i)$ & set of neighbours $j$ having an edge to vertex $i$ in $\mathcal{G}$ \\

\hdashline

$\bm{x}$ & spatial coordinates\\
$u_t(\bm{x})$  & the state, a random field indexed by both time $t$ and space $\bm{x}$ \\

$\bm{u}_t$ &
state vector at $t$ of state elements over discretised space\\

$\bm{u}= [\bm{u}_1,\ldots, \bm{u}_T ]^\top$ & state vector at discretised indexed times $1\leq t \leq T$ \\

$\bm{y}_t$ & response of map $\bm{y}_t = \bm{h}_t(\bm{u}_t)$ on state, often simply $\bm{h}(\bm{u}_t) = \bm{H}\bm{u}_t$ \\
$\bm{d}_t$ & noisy observation of response, $\bm{d}_t = \bm{y}_t + \bm{\epsilon}_t$, $\bm{\epsilon}_t\sim N(\bm{0}, \bm{\Sigma}_{\epsilon_t})$ at $t$ \\
$\bm{u}_{t|s}$ & state random variable $\bm{u}_t$ conditioned on observations $\bm{d}$ up until time $s$  \\
$\bm{u}|\bm{d}_T$, $\bm{u}|\bm{d}_t$, $\bm{\theta}|\bm{d}_t$ & smoothing, filtering, and parameter estimation problems \\

\hdashline 
$p(\bm{u})$ & probability density function of a random vector $\bm{u}$ \\
$\bm{u} \perp \bm{v}$ & independence of $\bm{u}$ and $\bm{v}$, $p(\bm{x},\bm{y}) = p(\bm{x})p(\bm{y})$ factorises \\

$\E{\bm{u}}$ & expectation operator $\E{\bm{u}}=\int p(\bm{u})~d\bm{u}$ \\
$\bm{\mu}$, $\bm{\Sigma}$ & mean $\bm{\mu}=\E{\bm{u}}$ and covariance $\bm{\Sigma}=\E{(\bm{u}-\bm{\mu})(\bm{u}-\bm{\mu})^\intercal}$\\
$\covsample$ & sample-covariance estimate \\
$\bm{\eta}$, $\bm{\Lambda}$ & canonical parametrisation $\bm{\eta}=\bm{\Lambda}\bm{\mu}$ and $\bm{\Lambda}=\bm{\Sigma}^{-1}$ \\


$\KLD\left(P||Q\right)$ & Kullback-Leibler divergence of (truth) $p_P(\bm{u})$ from (model) $p_Q(\bm{u})$ \\

\hdashline

$\bm{u}^{(i)}$ & realisation sampled from $\bm{u}^{(i)}\sim p(\bm{u})$ \\
$\bm{y}^{(i)}$, $\bm{d}^{(i)}$ & response, $\bm{y}^{(i)}=\bm{h}\left(\bm{u}^{(i)}\right)$, and noisy response, $\bm{d}^{(i)}=\bm{y}^{(i)} + \bm{\epsilon}^{(i)}$, realisations \\
$\bm{u}_{t|t-1}^{(i)}$, $\bm{u}_{t|t}^{(i)}$, $\bm{u}_{t|T}^{(i)}$ & prior, posterior filter and posterior smoothing realisation \\

$1 \leq i \leq n$ & realisation index (typically in parenthesised superscript $\bm{u}_t^{(i)}$ \\
$1 \leq (s,t) \leq T$ & time indexation, typically in subscripts  \\
$1 \leq k \leq p_{\bm{x}}$ & spatial position index of $\bm{x}$ in a discretised field \\
$1 \leq j \leq p$ & parameter index (dimensions of $\bm{u}$ or $\bm{u}_t$), enumerates combinations of $t$ and $k$ \\
$\mathcal{G}_{\bm{u}}, \mathcal{G}_{\bm{x}}$ & graphs defining relations between parameters and points in space \\

\hline
\end{tabular}
\label{table:nomenclature}
\end{table}

This section introduces the necessary background.
Detailed nomenclature is provided in Table \ref{table:nomenclature}.
Familiarity with the following three fields, with suggested references, are helpful:
\begin{itemize}
    \item Gaussian Markov random fields (GMRF) \citep{rue2005gaussian} and their relation to stochastic partial differential equations (SPDE) \citep{lindgren2011explicit, lindgren2022spde}.
    \item Classical ensemble based data assmilation \citep{evensen2009data} and Bayesian consistent assimilation through methods of triangular measure transport \citep{marzouk2016introduction, baptista2020representation, ramgraber2023ensemble1}.
    \item Supervised learning \citep{hastie2009elements} and the relation of statistical complexity to overfitting \citep{akaike1974new, claeskens2008model}.
\end{itemize}

\subsection{Dynamical model conditioned on observations}
\label{subsec:dynamical-model-conditioned}
Data assimilation (DA) consists of observations or data and a (dynamical forward) model.
As the forward model, we take a solution $g$ of the time-homogeneous stochastic process, $u_t(\bm{x})$ named the \textit{state}, defined via the SPDE
\begin{align}\label{eq:general-spde}
    \mathcal{L} u_t(\bm{x}) = \mathcal{A} u_t(\bm{x}) + \sigma(u_t(\bm{x})) \mathcal{W}_t(\bm{x}).
\end{align}
It consists of spatial and temporal model operators $\mathcal{A}$ and $\mathcal{L}$, and a (stochastic) diffusion term
$\sigma$ as $\mathcal{W}_t(\bm{x})$ is a white noise process.
This is a similar model as in the classical DA literature \citep[Chap.~4.3.2,~p.39]{evensen2009data}.
But, we are explicit that both $\mathcal{A}$ and $\mathcal{L}$ are local operators, and only work on a neighbourhood of $u_t(\bm{x})$ in time and space (e.g. differential operators or functions on $u_t(\bm{x})$ directly).
The operator $\mathcal{L}$ relates to derivatives in time.
We assume $\mathcal{W}_t(\bm{x})$ to be the differential of a spatio-temporal Wiener process, and take the integral to be the Ito integral, see e.g. \citet{oksendal2003stochastic}.
All three operators may depend on a set of parameters $\theta$ that may be taken as deterministic or stochastic.
Unless otherwise stated, assume $\bm{\theta}$ to be deterministic and their inclusion in $g$ implicit.

The second key component is (uncertain or noisily observed) data, $\bm{d}_t\sim p_{\bm{d}}$.
The observations are assumed related to the state through a (possibly stochastic) map $\bm{h}$.
Necessarily, the observations also yield information regarding $\bm{u}$.
The goal of DA is to retrieve information on the posterior of the state $\bm{u}|\bm{d}$.

An ensemble based data assimilation (EDA) approach simulates realisations $\bm{u}_{t|t-1}^{(i)}=\bm{g}\left(\bm{u}_{t-1|t-1}^{(i)}\right)$ through Monte Carlo sampling of the initial state $\bm{u}_{t-1|t-1}^{(i)}$ and typically numerical integration of Equation \eqref{eq:general-spde}.
The goal of EDA is then to update each realisation, mapping $\bm{u}_{t|t-1}^{(i)}$ to $\bm{u}_{t|t}^{(i)}$, so that $\bm{u}_{t|t}^{(i)}\sim p\left(\bm{u}_t|\bm{d}_t\right)$.

\begin{figure}
    \centering
    \includegraphics[width=1.0\linewidth]{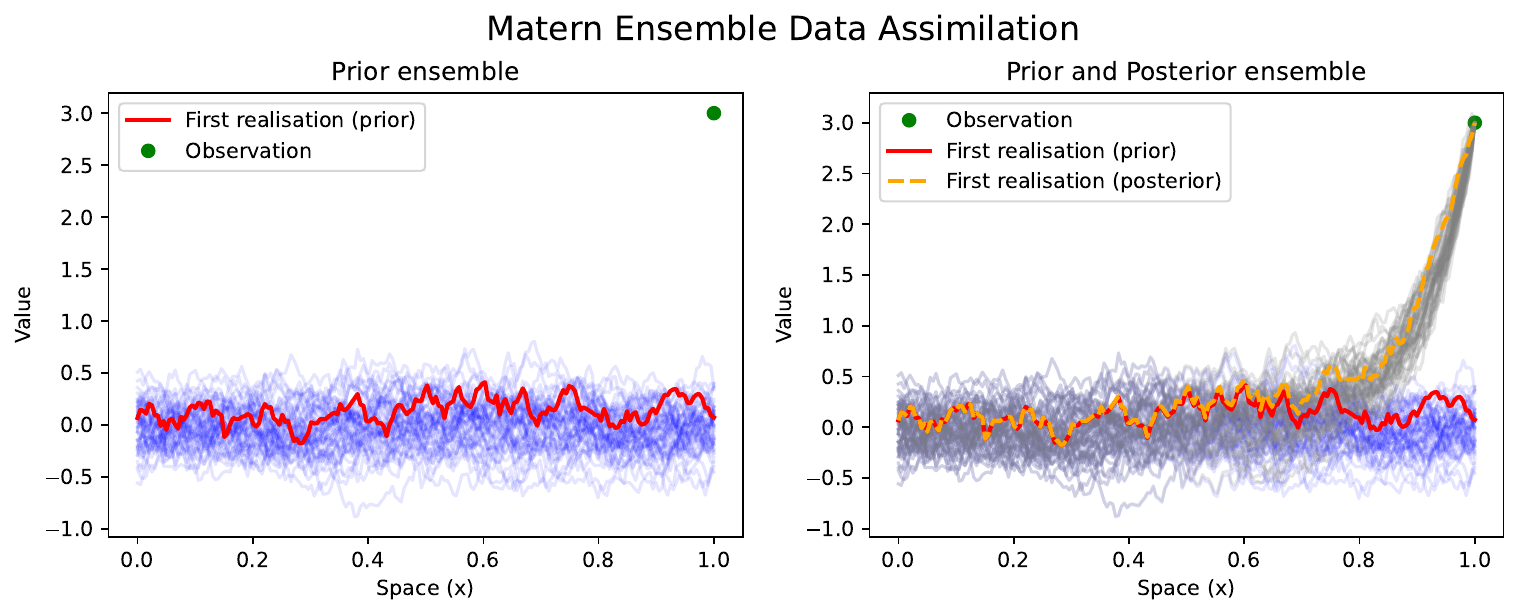}
    \caption{Left: 50 realisations of 1-d Matérn ($\kappa=0.1$) Gaussian process as in Example \ref{ex:conditional-matern-exact} sampled on $x\in[0,1]$, unconditioned on an observation of the endpoint (green dot) at $x=1.0$. Right: Conditioned realisations overlayd the unconditional ones from the left plot. The update is a special case of the EnKF and is expressed in Example \ref{ex:conditional-matern-exact}. Red line showcases the first Matérn realisation (left) and how this realisation is updated (orange, right).}
    \label{fig:matern-exact}
\end{figure}

\begin{example}[Conditioned Matérn-GRF]\label{ex:conditional-matern-exact}
The Matérn SPDE with $\nu=\frac{1}{2}$ and taken in one dimension is
\begin{align}\label{eq:matern-spde-1d}
    \left(\kappa^2 - \frac{d^2}{dx^2}\right)^{1/2} u(x) = \mathcal{W}(x).
\end{align}
The solution is a stationary mean-zero Gaussian process with covariance function
$\operatorname{Cov}(u_x, u_{x+h}) = \frac{\kappa}{2} e^{-\kappa^{-1}|h|}$.
It is Markov and has a mean-reverting property.
Assume we observe $d_{p}=u(p)+\epsilon$ a direct but noisy observation of $u(p)$ with $\epsilon \sim \mathcal{N}(0,\sigma_{\epsilon}^2)$.
Simulate $i=1,\ldots,n$ realisations of $u_{x|x_0}$ at $x=1,2,\ldots,p$, and add simulated noise $\epsilon^{(i)}$ to each $u(p)^{(i)}$ obtaining simulations of $d_p$.
Then update each realisation element (indexed by $x$) as
\begin{align}
    u_{x|p}^{(i)} = u_{x|0}^{(i)} - 
    \frac{\operatorname{Cov}(u_x,u_{p})}{\operatorname{Cov}(u_p,u_{p})+\sigma_{\epsilon}^2}
    \left(d_p^{(i)} - d_p\right)
    = 
    u_{x|0}^{(i)} - 
    \frac{
        \kappa e^{-\kappa^{-1}|p-x|}
    }{\kappa+2\sigma_{\epsilon}^2}
    \left(d_p^{(i)} - d_p\right)
\end{align}
yielding updated realisations $\bm{u}^{(i)}=[u_{1|p}^{(i)},\ldots,u_{p|p}^{(i)}]^\intercal \sim p\left(\bm{u}|\bm{d}_p\right)$.
\end{example}



Notice how in Example \ref{ex:conditional-matern-exact} we simulated both $\bm{u}$ and $\bm{d}$, and employed the joint Gaussian $P$ on $(\bm{u}, \bm{d})$.
In most situations the true distributions $P$ is unknown.
This is the reason for the ensemble approach, as realisations will be correctly sampled from $P$.
An approximation $Q$ to $P$ may then be estimated from the ensemble, and employed for the conditioning upon $\bm{d}$.


\subsection{Consistent ensemble based data assimilation}
\label{subsec:triangular-measure-transport}
Arguably, the most promising general purpose way to estimate $Q$ for $\bm{\nu}=[\bm{u},\bm{d}]^\intercal$ and update an ensemble is through Triangular Measure Transport \citep{marzouk2016introduction, baptista2020representation}.
Measure transport relies on the change of variable formulae
\begin{align}\label{eq:measure-transport--change-of-variable}
    p_{\bm{\nu}}(\bm{\nu}) = p_{\bm{z}}\left(\bm{S}(\bm{\nu})\right)\det \nabla \bm{S}(\bm{\nu})
\end{align}
to define the density $p_{\bm{\nu}}$ through a reference density $p_{\bm{z}}$ and a monotone map $\bm{S}$.
Triangular measure transport is obtained through structuring the map $\bm{S}$ as the \textit{Knothe-Rosenblatt} (KR) \textit{rearrangement} \citep{rosenblatt1952remarks}
\begin{align}\label{eq:knothe-rosenblatt-rearrangement}
    \bm{S}(\bm{\nu}) = 
    \left[
    \begin{array}{l}
        S_1(\nu_1)\\
        S_2(\nu_1,\nu_2)\\
        \vdots\\
        S_p(\nu_1,\ldots,\nu_p)
    \end{array}
    \right]
    =
    \left[
    \begin{array}{l}
         z_1  \\
         z_2 \\
         \vdots\\
         z_p
    \end{array}
    \right]
    = \bm{z}.
\end{align}
This choice essentially limits the space of possible transport maps to those that are monotone and bijective \citep{marzouk2016introduction}.
Note that the KR map, along with letting the reference $p_{\bm{z}}$ be the standard Gaussian, is the exact choice of recent research towards consistent data assimilation through triangular transport 
\citep{spantini2019coupling, baptista2020representation, baptista2021learning, ramgraber2023ensemble1}.

A Bayesian update of the prior ensemble can be done through the inverse KR map.
Characterise $\bm{d}$ through the change of variable formulae with a standard Gaussian reference and the KR map $\bm{S}_{\bm{d}}$.
The joint KR map becomes 
\begin{align}\label{eq:triangular-transport-joint-map}
    \bm{S}(\bm{d},\bm{u}) = 
    \left[
    \begin{array}{l}
        S_1(d_1)\\
        \vdots\\
        S_m(d_1,\ldots,d_m)\\
        \hdashline
        S_{m+1}(d_1,\ldots,d_m,u_1)\\
        \vdots\\
        S_{m+p}(d_1,\ldots,d_m,u_1,\ldots,u_p)
    \end{array}
    \right]
    =
    \left[
    \begin{array}{l}
         S_{\bm{d}}(\bm{d}) \\
         S_{\bm{u}}(\bm{d}, \bm{u})
    \end{array}
    \right]
    =
    \left[
    \begin{array}{l}
         z_1  \\
         \vdots\\
         z_{m+p}
    \end{array}
    \right]
    = \bm{z}
    .
\end{align}
Using the properties of the inverse, we may sample from the conditional $p(\bm{u}|\bm{d})$
through inversion of $\bm{u}\mapsto S_{\bm{u}}(\bm{d}, \bm{u})$ working with the lower block of the joint KR map.
We might either
\begin{enumerate}
    \item Sample $\bm{z}\sim N(0,\bm{I}_p)$ and evaluate $\bm{u}=S_{\bm{u}}^{-1}(\bm{d},\bm{z})\sim p(\bm{u}|\bm{d})$, which works if ${\bm{S}}$ is \textit{exact}.
    \item First sample $\tilde{\bm{z}}^{(i)}$ using a training sample / ensemble realisation $(\bm{u}_{t|t-1}^{(i)},\bm{d}_{t}^{(i)})$ evaluating $\tilde{\bm{z}}^{(i)}=S_{\bm{u}}(\bm{u}_{t|t-1}^{(i)},\bm{d}_{t}^{(i)})$.
    Then, sample from $p(\bm{u}|\bm{d}_{t})$ evaluating $\bm{u}_{t|t}^{(i)}=S_{\bm{u}}^{-1}(\bm{d}_{t},\tilde{\bm{z}}^{(i)})$. We here use subscript conditional notation to differentiate prior and posterior updated realisation.
\end{enumerate}
The second approach can be shown to work better because approximations in a learned map $\hat{\bm{S}}$ can be shown to cancel when writing the second step as a composite map
\begin{align}\label{eq:triangular-transport-composite}
    \bm{u}_{t|t}^{(i)} = S_{\bm{u}}^{-1}(\bm{d}_{t},\cdot) \circ S_{\bm{u}}(\bm{u}_{t|t-1}^{(i)},\bm{d}_{t}^{(i)}).
\end{align}

The KR choice allows learning $Q$ through marginal regressions $S_j$'s from separating the joint relative Kullback-Leibler Divergence (KLD) (see \citet{ramgraber2023ensemble1})
\begin{align}\label{eq:kullback-leibler-divergence}
    \KLD(P||Q) = E_P\left[\log p_P(u)\right] - E_P\left[\log p_Q(u)\right]
\end{align}
leading to minimisation of the following objective
\begin{align}\label{eq:triangular-transport-objective}
   \mathcal{J}_j(S_j) = E_{P}
   \left[
   \frac{1}{2}S_j(\nu_1,\ldots,\nu_j)^2 
   - \log \frac{\partial S_j(\nu_1,\ldots,\nu_j)}{\partial \nu_j}
   \right]
\end{align}
for each row $j$ in $\bm{S}$.
Here the choice of a standard Gaussian reference has been used.
The following example is from \citet[Appendix B]{ramgraber2023ensemble1}

\begin{example}
[Affine Triangular Measure Transport]
\label{ex:linear-triangular-measure-transport}
Let $\bm{S}$ be affine and choose the reference $p_{\bm{z}}$ as standard Gaussian.
Then
\begin{align}\label{eq:triangular-affine-map}
    \bm{S}(\bm{\nu}) = 
    \bm{C}\bm{\nu}
    =
    \left[
    \begin{array}{cccc}
        c_{1,1}\\
        c_{2,1} & c_{2,2} \\
        \vdots & \vdots & \ddots \\
        c_{p,1} & c_{p,2} & \ldots c_{p,p}
    \end{array}
    \right]
    \left[
    \begin{array}{c}
         \nu_1  \\
         \nu_2 \\
         \vdots\\
         \nu_p
    \end{array}
    \right]
\end{align}
yields $Q$ as multivariate Gaussian with precision $\bm{\Lambda}=\bm{C}^\top \bm{C}=\bm{\Sigma}^{-1}$.
Furthermore, the composite map \eqref{eq:triangular-transport-composite} retrieves Gaussian conditioning
\begin{align}\label{eq:knothe-rosenblatt-rearrangement-affine}
    \bm{u}_{t|t}^{(i)} = \bm{u}_{t|t-1}^{(i)} 
    - \bm{\Sigma}_{\bm{u},\bm{d}}
    \bm{\Sigma}_{\bm{d}}^{-1}
    \left(\bm{d}_{t|t-1}^{(i)} - \bm{d}_t\right).
\end{align}
Note how the update in Example \ref{ex:conditional-matern-exact} is a special case.
\end{example}

If Equation \eqref{eq:triangular-transport-objective} could be solved, then it is likely that $Q=P$ can be found.
If $P$ is non-Gaussian but $\bm{S}$ is constrained to be linear like $\bm{C}$ in Example \ref{ex:linear-triangular-measure-transport}, then the population estimate of covariance will be found by optimising \eqref{eq:triangular-transport-objective} and subsequently used in the update \eqref{eq:knothe-rosenblatt-rearrangement-affine}.
In practice, the idealised objective \eqref{eq:triangular-transport-objective} is exchanged with its sample estimator, using the ensemble.
This simple swap, if done naively and at large dimension $p+m$ of $\bm{\nu}$ at finite ensemble sizes $n$, has severe consequences like spurious correlations and ensemble collapse.


\subsection{Supervised learning of the statistical model $Q$}
\label{subsec:supservised-learning-of-statistical-model}

Because $P$ is unknown, and we use the KLD from Equation \eqref{eq:kullback-leibler-divergence}, we replace Equation \eqref{eq:triangular-transport-objective} with its sample estimator using the ensemble. Let $\bm{\theta}$ parameterise $Q$. Then, population quantities
\begin{align}
\label{eq:relative-kld-population-est}
\bm{\theta}_0 = \arg\min_{\theta}
-E_P\left[\log p_Q(\bm{u};\bm{\theta})\right]
\end{align}
are replaced with the likelihood estimator
\begin{align}
\label{eq:empirical-likelihood}
\hat{\bm{\theta}} = \arg\min_{\bm{\theta}}
-\frac{1}{n} \sum_{i=1}^n \log p_Q(\bm{u}^{(i)};\bm{\theta}).
\end{align}
Under regularity conditions, $\hat{\bm{\theta}}$ is asymptotically unbiased and is the minimum variance estimator (AUMVE) \citep{van2000asymptotic}. At finite sample/ensemble sizes, $\hat{\bm{\theta}}$ has non-zero variance. Thus, even if the model \eqref{eq:general-spde} dictates local behaviour and zero correlations for distant points, the estimated dependence in sample covariance (typically taken as $\bm{\theta}$ in EDA) will still be non-zero with probability one. This is known as spurious correlations in the literature.

\begin{example}[Ensemble Smoother on Matérn]
\label{ex:matern-ensemble-smoother}
Let $\bm{C}$ in Example \ref{ex:linear-triangular-measure-transport} be learned dense from Matérn data in Example \ref{ex:conditional-matern-exact}, optimising the empirical version \eqref{eq:empirical-likelihood} of \eqref{eq:triangular-transport-objective}.
For $p\geq n$ and otherwise under suitable conditions, we obtain a covariance estimate close to the sample covariance estimate.
Inserting the sample covariance into \eqref{eq:knothe-rosenblatt-rearrangement-affine}, we obtain an Ensemble Kalman Smoother \citep{evensen2009data}.
Figure \ref{fig:matern-ensemble-smoother} (left) showcase an update of the unconditioned Matérn realisations of Example \ref{fig:matern-exact} using the Ensemble Smoother. The right plot showcase the ensemble sample-correlations used for the update, along with additionally sampled sample-correlations and confidence bands that are functions of ensemble size.

\begin{figure}
    \centering
    \includegraphics[width=1.0\linewidth]{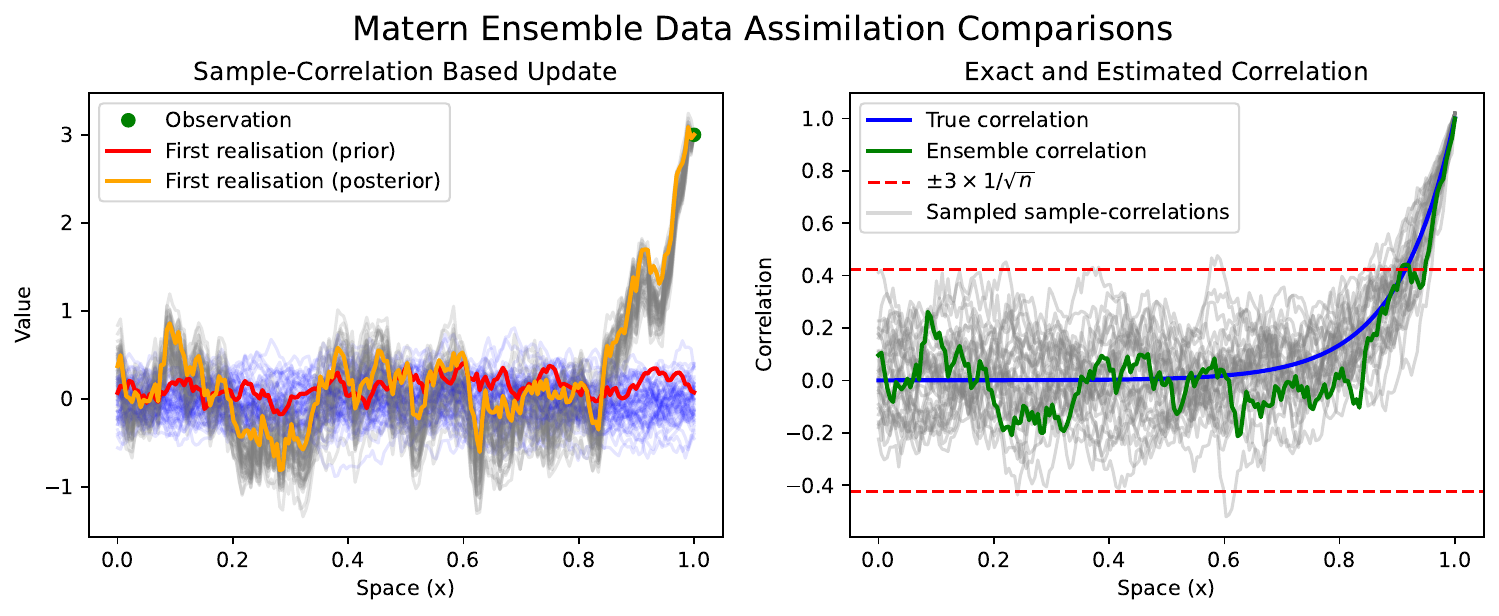}
    \caption{
    Left: Conditioned realisations as in Figure \ref{fig:matern-exact} right, but exchanging true covariance with sample covariance using the 50 unconditioned Matérn realisations.
    Right: 30 different sample correlations with the endpoint $x=1.0$ based on $n=50$ realisations, along with the true population correlation.
    The green line is the exact sample-correlations used for the update in the left plot.
    }
    \label{fig:matern-ensemble-smoother}
\end{figure}
\end{example}

Example \ref{ex:matern-ensemble-smoother} with Figure \ref{fig:matern-ensemble-smoother} illustrate how the estimated correlations far from the observation are non-zero and induce a noisy update.
The unfortunate effect fortunately goes to zero at an infinite ensemble size, as the variance of the statistical estimate is $\mathcal{O}(n^{-1})$ (middle plot).

A second unfortunate effect occurs after propagating a Bayesian update through the spurious correlations:
Doing this leads to an unwarranted decrease in uncertainty. 
It is a function of unwarranted (spurious) estimated strength of the dependence, and the number of times one propagates a Bayesian update, i.e. the number of observations one conditions on.
If (nearly) all uncertainty in one or multiple variables is lost to spurious updates, all ensemble members will share (near-identical) values in those variables. This loss of variation is known as ensemble collapse.
This effect, even with only one observation, is clear when comparing the width of the posterior ensembles of Examples \ref{fig:matern-exact} and \ref{fig:matern-ensemble-smoother}.

In the EDA literature, these statistical issues have received much attention. Two classes of solutions worth mentioning are:
\begin{itemize}
    \item \textbf{Distance-based localisation} \citep{houtekamer2001sequential, anderson2003local, hunt2007efficient} reduces the update to zero based on distance. A kernel function, such as the Gaussian kernel $\exp{-c\delta^2}$, is used, with $c$ controlling the radius of influence.
    \item \textbf{Adaptive localisation} involves regression $h^{-1}: \bm{d}\mapsto \bm{u}$ (with $\bm{K}$ as the asymptotic solution) and model selection. If the sample correlation between $(\bm{u}_j,\bm{d}_k)$ is smaller than a threshold (e.g. $3/\sqrt{n}$), the effect in $\bm{K}$ is set to zero. Otherwise, it is computed normally.
\end{itemize}

The KLD perspective \eqref{eq:kullback-leibler-divergence} provides more insight into EDA's statistical issues than just the variance of the sample covariance. When we exchange \eqref{eq:relative-kld-population-est} with \eqref{eq:empirical-likelihood}, we implicitly integrate over the empirical distribution function $P^*$:
\begin{align}\label{eq:negative-log-likelihood-empirical-distribution-relative-kld}
-\frac{1}{n}\sum_{i=1}^n \log p_Q(\bm{u}^{(i)};\bm{\theta})
= -E_{P^*}\left[\log p_Q(\bm{u};\bm{\theta})\right].
\end{align}
Since $Q(\hat{\bm{\theta}})$ is closer to $P^*$ than $P$ in the KLD sense, there is an expected bias $B(\hat{\bm{\theta}})>0$ when subtracted from the expectation under $P$ \citep{konishi1996generalised}. Furthermore, $B(\hat{\bm{\theta}})$ increases with the complexity of $\bm{\theta}$. If $\bm{\theta}$ has unnecessary degrees of freedom, it increases $B(\hat{\bm{\theta}})$ and the unwanted divergence from $Q$ to $P$.

Under regularity conditions, the bias $B(\hat{\bm{\theta}})$ may be approximated as:
\begin{align}
\label{eq:bias-relative-kld}
B(\hat{\bm{\theta}}) \approx
\tr\left(
\mathcal{H}(\bm{\theta}_0)
\operatorname{Cov}(\hat{\bm{\theta}})
\right),
\text{ where }
\mathcal{H}(\bm{\theta}_0) =
-E_P\left[
\nabla_{\bm{\theta}_0}^2
\log p_Q(\bm{u};\bm{\theta}_0)
\right].
\end{align}
Using the Sandwich estimator \citep{huber1967behavior} and empirical Hessian, we arrive at the TIC \citep{takeuchi1976distribution}. The expression for $B(\hat{\bm{\theta}})$ can also be related to effective degrees of freedom in $Q(\bm{\theta})$ \citep{hastie2009elements}. If $P=Q(\bm{\theta}_0)$, then $B(\hat{\bm{\theta}})=\operatorname{len}(\bm{\theta})$, recovering the celebrated AIC \citep{akaike1974new}.

This poses a problem for classical EDA: using the sample covariance, $\operatorname{len}(\bm{\theta})$ is quadratic in $p$, while the data-ensemble grows only linearly with increased resolution. Figure \ref{fig:ou-kld-vs-resolution} shows how the numerical scheme can improve with resolution while EDA using the sample covariance (EnKF/ES) only improves initially before statistical noise dominates through an increase in $B(\hat{\bm{\theta}})$.



Solutions to \eqref{eq:general-spde} are provided through numerical methods (e.g., finite element (FEM) or finite difference methods (FDM)) that improve at increased resolution; $\Delta \bm{x}$ and $\Delta t$ approaching zero. Fitting a high-resolution statistical model $Q(\hat{\bm{\theta}})$ close to $P$ or $Q(\bm{\theta}_0)$ for conditioning on $\bm{d}$ is non-trivial. If done naively, like with the sample covariance, the statistical model will diverge from $P$.
This is illustrated further in Section~\ref{subsec:convergence-to-the-true-solution}.

Fortunately, the KLD perspective provides insight into tools to limit divergence, as seen in Equation \eqref{eq:bias-relative-kld} for $B(\hat{\bm{\theta}})$. Our goal is the minimisation of the bias-corrected negative log-likelihood:
\begin{align}
\label{eq:bias-corrected-nll}
-E_{P}\left[\log p_Q(\bm{u};\bm{\theta})\right]
\approx
-E_{P^*}\left[\log p_Q(\bm{u};\bm{\theta})\right] + B(\hat{\bm{\theta}}).
\end{align}
To this end, we may:
\begin{itemize}
    \item \textbf{Use structure} to limit the size of $\bm{\theta}$ by constraining the search to a suitable subspace of the full parameter space.
    \item \textbf{Use regularisation} to trade variance in $\bm{\theta}$ for some bias. This makes the empirical negative log-likelihood larger but benefits from a smaller $B$, ultimately reducing the expression. \citet{hastie2009elements} discuss such techniques extensively.
\end{itemize}
It is important to encode both structure and regularisation before statistical fitting. This ensures that estimates $\hat{\bm{\theta}}$ have optimal properties and adapt to the simulated data in the ensemble.

Regularisation alone cannot solve the aforementioned resolution problem. At higher resolutions, statistical noise dominates (unless structure is informed upon), and optimal regularisation will zero out interactions (e.g., $\bm{K}=\bm{0}$), retaining the prior. Thus, structure is needed. Fortunately, the numerical solution of the SPDE model \eqref{eq:general-spde} provides structure in $P$.

\section{Markov structure from local operators}\label{sec:markov-structure-from-local-operators}

A key difference between ordinary supervised learning and EDA is the presence of the model in Equation~\eqref{eq:general-spde}, which induces structure in the target distribution.
In particular, locality in the underlying dynamics leads to conditional dependence structures that can be exploited in statistical modelling. 
Exact Markov properties need not hold in the target distribution $P$; it is sufficient that the dynamics exhibit a preference for locality, making sparse Markov structure a parsimonious and statistically efficient approximation under finite ensemble estimation.

\subsection{Locality and conditional independence}
\label{subsec:locality and conditional independence}

Locality properties often arise in $P$ from both analytical and numerical solutions to the model \eqref{eq:general-spde}.
Conditional independence provides a convenient way to encode (possibly approximately) this locality in $Q$.
In the Gaussian case, conditional independence is expressed through sparsity in the precision matrix. 
From \citet{rue2005gaussian}, if $\mathcal{G} = (\mathcal{V}, \mathcal{E})$ is a graph with vertices $\mathcal{V}$ and edges $\mathcal{E}$, a random vector $\bm{u} \in \mathbb{R}^d$ is a Gaussian Markov Random Field (GMRF) with respect to $\mathcal{G}$, mean $\bm{\mu}$, and SPD precision matrix $\bm{\Lambda} = \bm{\Sigma}^{-1}$, if
\begin{align}
p(\bm{u}) = (2\pi)^{-\frac{d}{2}} \sqrt{|\bm{\Lambda}|} \exp \left( -\frac{1}{2} (\bm{u} - \bm{\mu})^\intercal \bm{\Lambda} (\bm{u} - \bm{\mu}) \right),
\end{align}
and
\begin{align}
\Lambda_{i,j} \neq 0 \;\Leftrightarrow\; (i,j) \in \mathcal{E}, \quad i \neq j.
\end{align}

Thus, conditional independence is encoded directly through zeros in $\bm{\Lambda}$, restricting the parameter space to SPD matrices consistent with $\mathcal{G}$. 
The number of statistical parameters is then $m + l$, where $m = |\mathcal{V}|$ and $l = |\mathcal{E}|$, compared to $O(m^2)$ parameters for a dense covariance model. 
This reduction is critical in ensemble data assimilation, where the sample covariance corresponds to a fully connected graph and leads to overfitting under finite ensemble sizes.

In many dynamical systems, the governing equations are local, so that each state variable interacts primarily with nearby variables in space and time. 
This induces a preference for local dependence structures, where the dominant conditional dependencies are captured by a sparse graph $\mathcal{G}$. 
Imposing such structure in $Q$ therefore yields a parsimonious statistical model aligned with the underlying dynamics, while avoiding the estimation of spurious long-range dependencies.

The following sections make the connection between local models and conditional independence more precise.

\subsection{SPDE models and Markov structure}
\label{subsec:spde models and markov structure}

In terms of continuous solutions to \eqref{eq:general-spde}, \citet{rozanov1977markov} finds that for linear SPDEs of the form
\begin{align}\label{eq:linear-spde}
    \mathcal{L} u(\bm{x}) = \mathcal{W}(\bm{x}),
\end{align}
the solutions are Gaussian random fields.
When the operator $\mathcal{L}$ is local, in the sense that it involves only differential operators of finite order, the resulting field exhibits conditional independence structure. 
In particular, the precision operator $\mathcal{Q}$ is given by
\begin{align}
\mathcal{Q} = \mathcal{L}^* \mathcal{L},
\end{align}
where $\mathcal{L}^*$ denotes the adjoint of $\mathcal{L}$. 
The locality of $\mathcal{L}$ therefore carries over to $\mathcal{Q}$, implying that conditional dependencies are governed by local interactions. 
This provides a direct link between locality in the governing equations and sparse precision structure in the associated statistical model.


\begin{example}
[Markov solutions to the Matérn SPDE]
\label{ex:markov-solutions-to-matern-spde}
The general Matérn SPDE is given by 
\begin{align}
    \left(
    \kappa^2 - \laplacian
    \right)^\frac{\alpha}{2} \bm{u}(\bm{x})
    = \mathcal{W}(\bm{x}),
\end{align}
    which is a fractional linear SPDE.
    The solution is Gaussian, and in the case of integer powers of $\alpha$, the precision operator $\left(
    \kappa^2 - \laplacian
    \right)^{\alpha}$ can be expanded into a sum of integer powers of the negated Laplacian \citep{lindgren2022spde}.
    Thus $\mathcal{L}$ is linear and involves only ordinary local operators and is therefore also Markov.
    In the non-integer case, the negative fractional Laplacian is given by
    \begin{align}\label{eq:fractional-laplacian}
    (-\laplacian)^{\frac{\alpha}{2}} \bm{u}(\bm{x}) = C_{\alpha,d} \int_{\mathbb{R}^d} \frac{\bm{u}(\bm{x}) - \bm{u}(\bm{y})}{|\bm{x} - \bm{y}|^{d + \alpha}} \, d\bm{y},
\end{align}
    and is a global/non-local operator. The solution is still Gaussian, but not Markov.
    Due to the "kernel" $\frac{1}{|\bm{x} - \bm{y}|^{d + \alpha}}$, there is, however, still an importance to locality, as the influence of distant points $\bm{y}$ decreases with distance.
\end{example}

Recent work such as \citet{schafer2021sparse} emphasizes the screening effect from local operators, whereby conditional dependencies between distant variables become negligible when conditioning on nearby variables.
This provides a theoretical justification for sparse precision approximations in $Q$, even when exact Markov properties do not hold in $P$.

In EDA the target distribution $P$ is not defined through the exact solution of \eqref{eq:general-spde}, but rather the discretised numerical simulation.
But here, too, conditional independence arise.
Specifically, we will see that 
\begin{itemize}
    \item For EDA smoothing problems, a theoretical argument can (often) be made for exact Markov properties.
    \item For EDA filtering problems, a theoretical argument can be made for a Markov approximation, if the argument for the smoothing problem is valid.
    \item For EDA parameter estimation, a separate argument can be made for Markov properties and sparsity.
\end{itemize}

\subsection{Discretisation and sparse precision}
\label{subsec:discretisation and sparse precision}

The connection between locality and conditional independence carries over to numerical discretisations of Equation~\eqref{eq:general-spde}. 
In finite difference methods, local (differential) operators are approximated using neighbouring grid points, leading to band-sparse matrices. 
For instance in the case of Equation~\eqref{eq:linear-spde}, If $\tilde{\bm{L}}$ denotes a discrete approximation of $\mathcal{L}$, the resulting precision matrix
\begin{align}
\bm{\Lambda} = \tilde{\bm{L}}^\top \tilde{\bm{L}}
\end{align}
inherits this sparsity, implying conditional independence between distant states.

The celebrated SPDE approach of \citet{lindgren2011explicit} provides a general framework for constructing such models using finite element methods (FEM). 
Here, the solution is represented in terms of local basis functions, leading to sparse stiffness and (approximated diagonal) mass matrices, and consequently sparse precision matrices. Example~\ref{ex:fem solution to matern alpha 2 spde approach} is a classical illustration on the Matérn $\alpha=2$ case.

\begin{example}[FEM solution to Matérn ($\alpha=2$)]
\label{ex:fem solution to matern alpha 2 spde approach}
    For the Matérn $\alpha=2$ we have the SPDE 
    \begin{align}
    \left(
    \kappa^2 - \laplacian
    \right) \bm{u}(\bm{x})
    = \mathcal{W}(\bm{x})
\end{align}
Using a Galerkin FEM discretisation with local basis functions $\phi_i$, we obtain an approximately diagonal mass matrix and a sparse stiffness matrix
\begin{align}
    \tilde{M}_{ii} &= \sum_{j\in \neighbours(i)} \int_{\Omega_{ij}} \phi_i(\bm{x})\phi_j(\bm{x}) \ d\bm{x}, \\
    A_{ij}  &= \int_{\Omega_{ij}} \nabla \phi_i(\bm{x}) \nabla \phi_j(\bm{x}) \ d\bm{x},
\end{align}
where $\Omega_{ij}$ represents the union of the mesh elements where the basis functions $\phi_i$ and $\phi_j$ overlap.
We may then express the precision matrix as:
\begin{align}
  \bm{\Lambda}_{\bm{u}} = 
  \left(\kappa^2\tilde{\bm{M}} + \bm{A}\right)^\top 
  \tilde{\bm{M}}^{-1}  
  \left(\kappa^2\tilde{\bm{M}} + \bm{A}\right).
\end{align}
The sparsity of \(\bm{\Lambda}_{\bm{u}}\), arising from the locality of the basis functions, induces a conditional independence structure consistent with a GMRF.

\end{example}

Appendix~\ref{app:the-spde-approach} further illustrates the SPDE approach for a simplified version of Equation~\eqref{eq:general-spde} in Equation~\eqref{eq:general-spde-first-order-in-time}, leading to an explicit and sparse precision matrix.
For a more recent and broader overview of the approach, we refer to \citet{lindgren2022spde}.


Complementary to the FDM and SPDE approaches, the screening perspective in \citet{schafer2021sparse} shows that locality of the underlying operator also leads to efficient sparse representations of the precision matrix after discretisation via sparse Cholesky factorizations.

These results show that sparse conditional independence structure in the discretised setting, encoded through precision matrices, arises naturally from locality in the underlying continuous operators, providing a direct probabilistic and computational justification for imposing sparsity in $Q$.

\subsection{Implications for ensemble data assimilation}
\label{subsec:implications-for-eda}

The results above provide a principled justification for imposing conditional independence structure in ensemble data assimilation. 
From the statistical learning perspective in Section~\ref{sec:ensemble-based-data-assimilation}, the model $Q$ is estimated from a finite ensemble and therefore subject to overfitting. 
This induces a preference for parsimonious models, where unnecessary dependencies are avoided. 
Locality in the underlying dynamics provides a natural structure for such parsimony, as it implies that the dominant dependencies are local, while long-range interactions are weaker and harder to estimate reliably.

\paragraph{Smoothing.}
For smoothing problems, the full state trajectory $\bm{u}_{0:T} = (\bm{u}_0, \ldots, \bm{u}_T)$ is considered jointly. 
In many dynamical systems, the evolution is local in space and time, implying that $\bm{u}_t$ depends primarily on nearby states. 
This leads to exact or near-exact Markov structure in the joint distribution, which can be encoded through a sparse precision matrix. 
This is consistent with the SPDE formulation, where local operators induce sparse precision structure in space-time.

\paragraph{Filtering.}
For filtering problems, the target distribution is typically a marginal $p(\bm{u}_t \mid \bm{y}_{1:t})$, obtained by integrating out past states:
\begin{align}
p(\bm{u}_t \mid \bm{y}_{1:t}) = \int p(\bm{u}_{0:t} \mid \bm{y}_{1:t}) \, d\bm{u}_{0:t-1}.
\end{align}
This marginalisation generally destroys exact Markov properties that may exist in the corresponding smoothing problem, when the underlying dynamics are local. 
However, locality is often preserved approximately: conditional dependencies between distant variables become negligible when conditioning on nearby states, corresponding to a screening effect \citep{schafer2021sparse}. 

From the statistical learning perspective in Section~\ref{sec:ensemble-based-data-assimilation}, this provides a justification for sparse Markov approximations in $Q$. 
In particular, a $k$-order Markov approximation may provide a near-optimal trade-off between bias and variance when evaluated using the bias-corrected negative log-likelihood in Equation~\eqref{eq:bias-corrected-nll}. 
Under finite ensemble sizes, restricting the model to local dependencies reduces estimation variance, while increasing the Markov order $k$ reduces approximation bias. 
This suggests that the optimal Markov order $k$ depends on the ensemble size $n$, increasing as more data becomes available. 
This behavior is illustrated for the Lorenz-96 filtering problem in Section~\ref{subsec:filtering-lorenz-96} and Figure~\ref{fig:lorenz-96-markov-order}, where performance is profiled over $k$ for varying $n$.

\paragraph{Parameter estimation.}
For parameter estimation, the relationship between parameters, states, and observations is often complex and not explicitly known. 
Nevertheless, locality considerations may still be used to impose structure in the corresponding statistical model, for instance through sparse regression mappings between states and observations. 
This provides both regularisation and interpretability, as the resulting model highlights the dominant dependencies in the system.

Taken together, these considerations motivate modelling $Q$ with sparse precision structure, and learning sparse mappings between states, parameters, and observations. 
In the next section, we show how these ideas can be incorporated directly into ensemble-based data assimilation through the Ensemble Information Filter.

\section{An ensemble information filter: Incorporating Markov structure and regularisation in EDA}
\label{sec:an-ensemble-information-filter-encoding-Markov}

Common to all the scenarios described in Section \ref{sec:markov-structure-from-local-operators} is the importance of conditional independence as a meaningful modelling choice encoded in the model $Q$.
This conditional independence can be represented by a graph $\mathcal{G}$.
We now develop EDA that estimate prior distributions with respect to $\mathcal{G}$, enabling easy conditioning on observations.

It is worth noting that the EnKF formulation is not naturally suited for models with conditional independence. The EnKF framework employs a moment parameterisation of the Gaussian, either through the sample covariance matrix or its square root. 
To address conditional independence in this parametrisation requires dense matrix inversion, making it difficult to directly encode such properties into the covariance structure.

Instead of encoding this structure in the covariance, we encode conditional independence in the precision $\bm{\Lambda}_{\bm{u}}$ (specifically) and the KR map \( \bm{S} \) (generally). This encoding of structure aligns with the objective in Equation \eqref{eq:bias-corrected-nll}.
The complexity and non-linearity of \( \bm{S} \), as well as the need for appropriate regularisation, remain open questions. 
We will further assume that \( \bm{S} \) is of the affine type, which removes the question of non-linearity and significantly reduces the need for regularisation.

This assumption brings us back to the original goal of this paper: to develop a competitor to the highly successful EnKF. 
The new method should maintain the computational efficiency and scalability of the EnKF while eliminating the need for manual tuning, such as adaptive or distance-based localisation. 
However, we believe that tuning the non-linear complexity of \( \bm{S} \) to the specific problem or data is the next crucial step. 
Achieving this may lead to fully non-linear, non-Gaussian EDA computations at scale and that is truly adaptive to the problem.

\subsection{Encoding conditional independence from $\mathcal{G}$ in the KR-map $S$}
\label{subsec:encoding-conditional-independence-from-G-inthe-KR-map-S}

Conditional independence can be built into $\bm{S}$ by considering the inverse KR map
\begin{align}\label{eq:knothe-rosenblatt-rearrangement-inverse}
    \bm{S}^{-1}(\bm{u}) = 
    \left[
    \begin{array}{l}
        S_1^{-1}(z_1)\\
        S_2^{-1}(u_1;z_2)\\
        \vdots\\
        S_p^{-1}(u_1,\ldots,u_{p-1};z_p)
    \end{array}
    \right]
    =
    \left[
    \begin{array}{l}
         u_1  \\
         u_2 \\
         \vdots\\
         u_p
    \end{array}
    \right]
    = \bm{u},
\end{align}
which is evaluated from top to bottom \citep{ramgraber2023ensemble1, ramgraber2023ensemble2}.
Consider conditional independence w.r.t. the AR-1 graph for $p=4$ (see Figure \ref{fig:auto-regressive-1}), i.e. $u_3\perp u_1 | u_2$ and $u_4\perp u_1, u_2 | u_3$. Then, if the reference $p_{\bm{z}}$ factorises to a product distribution \citep{spantini2018inference},
writing $\argnotused{u_j}$ for redundant arguments and $\argused{u_j}$ for necessary, we get
\begin{align}
    \bm{S}^{-1}(\bm{u}) = 
    \left[
    \begin{array}{l}
        S_1^{-1}(z_1)\\
        S_2^{-1}(\argused{u_1}; z_2)\\
        S_3^{-1}(\argnotused{u_1}, \argused{u_2}; z_3) \\
        S_4^{-1}(\argnotused{u_1}, \argnotused{u_2}, \argused{u_3}; z_4)
    \end{array}
    \right]
    =
    \left[
    \begin{array}{l}
         u_1  \\
         u_2 \\
         u_3 \\
         u_4
    \end{array}
    \right]
    \text{ implies }
    \begin{array}{l}
         p(u_1)  \\
         p(u_2|\argused{u_1}) \\         
         p(u_3|\argnotused{u_1},\argused{u_2}) \\
         p(u_4|\argnotused{u_1},\argnotused{u_2},\argused{u_3})  
    \end{array}
    =
    \begin{array}{l}
         p(u_1)  \\
         p(u_2|\argused{u_1}) \\         
         p(u_3|\argused{u_2}) \\
         p(u_4|\argused{u_3})
    \end{array}.
\end{align}
Note that the conditional independence properties mean that $u_j$ is a function only of its neighbours, conditioned on the neighbours being available.
The matter of availability implies that the ordering of $\bm{S}$ w.r.t. $\bm{u}$ is important.
It turns out that for telescoping joint distributions, e.g. resulting from Markov properties in time, 
the optimal ordering is straightforward to find
\citep{ramgraber2023ensemble2}.
However, for a general graph $\mathcal{G}$, the question of ordering is more delicate.


Indeed, finding an optimal permutation \( \pi \) that maximises sparsity in the KR-map \( \bm{S} \) given the conditional independence graph \( \mathcal{G} \), is known to be an NP-hard problem. However, this is a well-researched area, and several algorithms that yield good approximate permutations are available \citep{cuthill1969reducing, amestoy2004algorithm, karypis1997metis}. 
Recent work by \citet{schafer2021sparse} develops an efficient approach for jointly determining a permutation and a sparse precision approximation, based on locality assumptions on the underlying operator. 
This provides a data- and model-driven alternative to purely graph-theoretic ordering strategies, and is closely related to the locality assumptions discussed in Section~\ref{sec:markov-structure-from-local-operators}.

Let \( \bm{S}_\pi \) denote the KR-map arranged according to a permutation \( \pi \) (with permutation matrix \( \bm{\Pi} \)), and let \( \pi_\mathcal{G} \) represent an optimal ordering that minimises the fill-in in the Cholesky decomposition \( \bm{L} \) of the precision matrix, where \( \bm{L}\bm{L}^\top = \bm{\Pi}_\mathcal{G}\bm{\Lambda}\bm{\Pi}_\mathcal{G}^\top \). As detailed in Appendix \ref{app:in-fill-reduction-on-CTC}, by applying the composition of the reverse permutation and the Cholesky-optimised permutation, \( \pi_\mathcal{G} \), we can derive the KR-map \( \bm{S}_{\pi_\mathcal{G}} \) with significantly reduced sparsity.

This approach ensures that the KR-map \( \bm{S}_{\pi_\mathcal{G}} \) inherits the sparsity benefits of the Cholesky factor \( \bm{L} \), allowing us to maintain computational efficiency while encoding most of the conditional independence structure of the model.

\subsection{The Ensemble Information Filter}
\label{subsec:the-ensemble-information-filter}

When the transformation \( \bm{S} \) in the KR-map is affine and the reference measure is Gaussian, the resulting distribution \( Q \) is also Gaussian. In this case, Gaussian conditioning is most naturally performed in the canonical parametrisation. The Information Filter (IF) \citep{kalman1961new, moore1979optimal} update equations are well-suited for this, as they operate directly on the precision matrix \( \bm{\Lambda} \) and the canonical mean \( \bm{\eta} = \bm{\Lambda} \bm{\mu} \). The update equations are:
\begin{align}\label{eq:information-filter-update}
    \bm{\eta}_{t|t} &= \bm{\eta}_{t|t-1} + \bm{H}^\top \bm{\Lambda}_{\bm{\epsilon}_t} \bm{y}_t,  \\
    \bm{\Lambda}_{t|t} &= \bm{\Lambda}_{t|t-1} + \bm{H}^\top \bm{\Lambda}_{\bm{\epsilon}_t} \bm{H},
\end{align}
where \( \bm{\Lambda}_{\bm{\epsilon}_t} \) is the precision of the observation noise \( \bm{\epsilon}_t \). While these equations are mathematically equivalent to the Kalman Filter (KF) \citep{kalman1960new} equations, the ensemble approach benefits from the precision matrix because it allows encoding conditional independence structures, thereby minimising divergence bias \( \bm{B} \) when \( \bm{\Lambda} \) is estimated from the ensemble.

To implement the IF equations \eqref{eq:information-filter-update} in an Ensemble Information Filter (EnIF), we must adapt them for use with ensemble realisations, which are not explicitly part of the standard equations. Define the canonical realisation as 
\begin{align}
    \bm{\eta}_{t|t-1}^{(i)} = \Lambda_{t|t-1} \bm{u}_{t|t-1}^{(i)},
\end{align}
the adjusted sampled observation as 
\begin{align}
    \tilde{\bm{d}}_t^{(i)}=\bm{d}-\bm{r}^{(i)},
\end{align}
where $\bm{r}^{(i)}$ is the noisy residual defined as 
\begin{align}
    \bm{r}^{(i)}=\bm{h}(\bm{u}_{t|t-1}^{(i)})-\bm{H}\bm{u}_{t|t-1}^{(i)}+\bm{\epsilon},
\end{align}
where \( \bm{\epsilon} \) is zero-mean Gaussian noise, sampled according to a precision matrix specifying observation uncertainty \( \bm{\Lambda}_{\bm{\epsilon}_t} \).

Appendix \ref{app:enif-enkf-equivalence} demonstrates that the following three steps are equivalent to the EnKF update as described in Example \ref{ex:conditional-matern-exact} and Example \ref{ex:linear-triangular-measure-transport}:
\begin{enumerate}
    \item Map realisations to canonical realisations: 
    \begin{align}
        \bm{\eta}_{t|t-1}^{(i)} = \Lambda_{t|t-1} \bm{u}_{t|t-1}^{(i)}.
    \end{align}
    \item Update each canonical realisation using the IF equations, along with an update of the precision:
    \begin{align}
        \bm{\eta}_{t|t}^{(i)} &= \bm{\eta}_{t|t-1}^{(i)} + \bm{H}^\top \bm{\Lambda}_{\bm{r}_t} \tilde{\bm{d}}_t^{(i)},  \\
        \bm{\Lambda}_{t|t} &= \bm{\Lambda}_{t|t-1} + \bm{H}^\top \bm{\Lambda}_{\bm{r}_t} \bm{H}.
    \end{align}
    \item Reverse-map each updated canonical realisation to an ordinary updated realisation: 
    \begin{align}
        \bm{u}_{t|t}^{(i)} = \Lambda_{t|t}^{-1} \bm{\eta}_{t|t}^{(i)}.
    \end{align}
\end{enumerate}

Thus, if \( \bm{\Lambda}_{t|t-1} \) is estimated from the ensemble of \( \bm{u}_{t|t-1} \) with respect to the graph $\mathcal{G}$ -- which encodes local (conditional independence) structure that is either generated or parsimoniously approximated from the solution to Equation \eqref{eq:general-spde} -- then the posterior ensemble will suffer significantly less from statistical estimation error, spurious correlations, and ensemble collapse compared to standard EnKFs. 
This is contingent on \( \bm{H} \) being either known or estimated with minimal statistical error.

Some important considerations for the EnIF are:

    \paragraph{Encoding conditional independence.} The canonical parametrisation is optimal for encoding conditional independence. Section \ref{sec:markov-structure-from-local-operators} shows how such structures can follow, at least approximately, from a local model \eqref{eq:general-spde}.
    
    \paragraph{Precision estimation.} For both statistical and computational scalability, \( \bm{\Lambda}_{t|t-1} \) must be efficiently estimated. A well-conditioned SPD result is obtained by estimating \( \bm{\Lambda} \) through the corresponding affine KR map \( \bm{C} \). Only the non-zero elements of a permutation-optimised \( \bm{C}_{\pi_\mathcal{G}} \) should be estimated. This method leverages the row-separation property, making it relatively scalable. However, not all sparsity is retained, and unnecessary degrees of freedom are often introduced.

    \paragraph{Locality is intrinsic: adaptive, automatic, and optimal.} Locality is explicitly encoded through estimation with respect to \( \mathcal{G} \).
    Consequently, the bias \( \bm{B} \) is expected to be relatively small, eliminating the need for ad-hoc localisation remedies. Since locality is encoded in the parametrisation before estimation, this approach is likely more efficient than traditional post-processing with localising kernels. Furthermore, it will adapt fully to the strength of dependence in the data, requiring no tuning of the "radius of convergence".

    \paragraph{Sparse observation operators $\bm{H}$.} The estimation of \( \bm{H} \) also matters for EnIF's scalability, since it often falls into the \( p >> n \) category. Regularisation is necessary to limit bias \( \bm{B} \). Moreover, to maintain sparsity in the posterior precision \( \bm{\Lambda}_{t|t} \), the estimation must produce a sparse result. The importance of $\bm{H}$ warrants the discussion in Section~\ref{subsec:learning-sparse-H}.
    

    \paragraph{EnKF equivalence under dense estimates.} Estimating $\bm{\Lambda}$ using an optimally poor ordering (obtaining a dense result $\bm{C}$ and $\bm{\Lambda}$), while taking $\bm{H}$ as the LLS estimator, the EnKF is obtained statistically speaking, but in a computationally inefficient manner.
    
    \paragraph{Statistical efficiency of separating $\bm{\Lambda}$ and $\bm{H}$ estimation.} TMT provides a full likelihood estimation approach for estimating \( Q \). This is done implicitly by placing \( \bm{d} \) first in the block structure of \( \bm{S}_{\bm{d},\bm{u}} \). In EnIF, \( Q \) is estimated using a combination of likelihood and regression. This necessitates an efficiency discussion, as divergence bias \( \bm{B} \) depends on the variance of the estimators. The AUMVE property of the full likelihood approach is partially retained. Furthermore, while \( \bm{y} \mid \bm{u} \) may not be Gaussian, it can be assumed uncorrelated since \( \bm{y} = g(\bm{u}) \) is derived from the governing equation. This yields properties close to Gauss-Markov and BLUE, indicating that the regression will be statistically efficient as well.

    \paragraph{Smoothing through extending the graph.} Although EnIF is presented as a filtering solution, the smoothing problem can also be addressed by allowing \( \bm{u}_{t} \) to include values of the state at times \( s \leq t \), with Markov properties in time encoded through \( \mathcal{G} \). For each Monte Carlo simulation, the state vector is appended with new values at time \( t \), as is \( \mathcal{G} \).

\paragraph{Extension to nonlinear observation operators.}
The formulation above assumes a linear or locally linear observation operator through \( \bm{H} \). In practice, many applications involve nonlinear mappings \( \bm{h} \). A standard approach in ensemble methods is to apply multiple data assimilation (MDA), as in ES-MDA \citep{emerick2013ensemble}, where the update is performed iteratively over several steps.

The EnIF framework admits an analogous extension, referred to as EnIF-MDA, where the canonical update is applied sequentially with appropriately inflated observation noise. In the linear case, this procedure is equivalent to the standard EnKF/ES-MDA formulation, as shown in Appendix~\ref{app:multiple-data-assimilation extension}. The canonical parametrisation simplifies the corresponding derivations.

\subsection{Learning sparse observation operators}
\label{subsec:learning-sparse-H}

In many applications, the observation operator $\bm{H}$ is not known explicitly, but must be estimated from ensemble data. 
This is typically a high-dimensional problem, where the state dimension $p$ is large relative to the ensemble size $n$, making the estimation of $\bm{H}$ highly susceptible to overfitting.

From a population perspective, the optimal linear observation operator corresponds to the expected gradient $\mathbb{E}[\nabla \bm{h}(\bm{u})]$ \citep{raanes2023review}. 
Thus, estimating $\bm{H}$ may be viewed as approximating a conditional expectation of the form
\begin{align}
\mathbb{E}[\bm{y}_t \mid \bm{u}_t] \approx \bm{H} \bm{u}_t.
\end{align}
Under finite ensemble sizes, this becomes a statistical learning problem, analogous to the estimation of the precision matrix.

Locality considerations, as developed in Section~\ref{sec:markov-structure-from-local-operators}, provide a natural structural assumption also for $\bm{H}$. 
Specifically, each observation is expected to depend only on a subset of the state variables, leading to a sparse regression structure. 
Imposing sparsity reduces estimation variance and avoids spurious long-range dependencies.

Sparsity in $\bm{H}$ is also important for computational reasons. 
From the update equation
\begin{align}
\bm{\Lambda}_{t|t} = \bm{\Lambda}_{t|t-1} + \bm{H}^\top \bm{\Lambda}_{\bm{\epsilon}_t} \bm{H},
\end{align}
it follows that dense estimates of $\bm{H}$ will generally destroy sparsity in the posterior precision matrix. 
Thus, sparse estimation of $\bm{H}$ is necessary to preserve the conditional independence structure of the model.
In this instance it becomes important for computational and not statistical reasons.

In practice, sparse estimation can be achieved using regularised regression methods. 
While the LASSO \citep{tibshirani1996regression} is a natural baseline, it may be computationally prohibitive in high-dimensional settings. 
Instead, we employ a boosting-based monotone LASSO \citep{hastie2007forward} with information-theoretic early stopping, as described in Appendix~\ref{app:fast-regularised-sparse-regression-through-boosting} and detailed in Algorithm~\ref{alg:monotone-lasso-with-early-stopping}.
This approach yields fast (see Figure~\ref{fig:lasso-monotone-timing}), adaptive, and highly sparse estimates of $\bm{H}$, suitable for large-scale applications.

Thus, sparsity in $\bm{H}$ complements sparsity in $\bm{\Lambda}$, yielding a coherent framework in which both prior structure and observation relationships are learned adaptively under locality constraints and finite-sample considerations.

\section{Synthetic numerical experiments}
\label{sec:synthetic-numerical-experiments}

We now conduct numerical experiments to illustrate the properties of the EnIF. The statistical objective of EDA, as defined in Equation
\eqref{eq:kullback-leibler-divergence},
is challenging to evaluate for high-dimensional distributions. However, by selecting examples where both the true distribution 
$P$ and the approximated distribution 
$Q$ are multivariate Gaussians, we can use the exact analytical expression for the KLD:
\begin{align}\label{eq:kullback-leibler-gaussian}
    \KLD \left(P||Q\right) = 
    \frac{1}{2}
    \left[
    \left(\mu_Q-\mu_P\right)^\intercal \Sigma^{-1} (\mu_Q-\mu_P) + 
    \tr \left(\Sigma_Q^{-1}\Sigma_P\right) -
    \log \frac{|\Sigma_P|}{|\Sigma_Q|} - 
    \operatorname{len}(\mu)
    \right].
\end{align}
This enables us to evaluate the performance of EnIF under the statistical challenges discussed in Section \ref{sec:ensemble-based-data-assimilation}.

\subsection{Convergence to the true solution in resolution}
\label{subsec:convergence-to-the-true-solution}

\begin{figure}[t]
    \centering
    \includegraphics[width=0.6\textwidth]{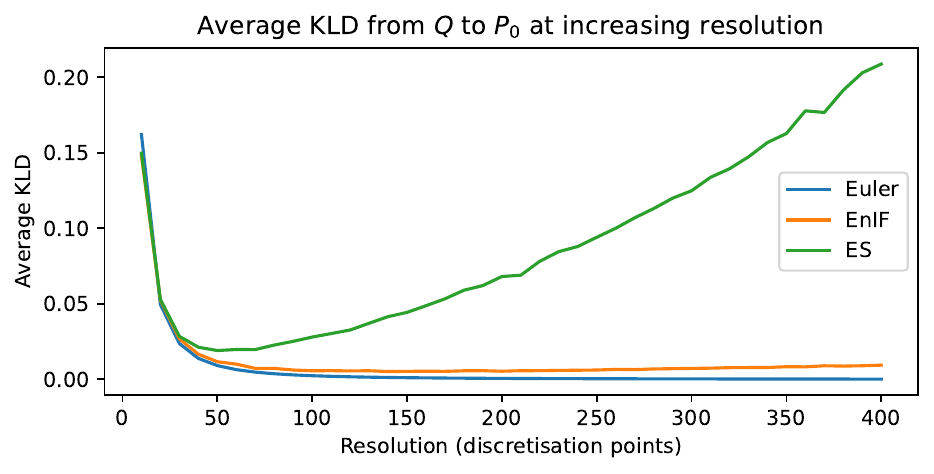}
    \caption{
    Average KLD values \eqref{eq:kullback-leibler-gaussian} for the EnIF (orange), EnKF/ES (green), and the Euler-Maruyama scheme (blue) relative to the true analytical posterior of the 1D Matérn process ($\kappa=1.0$) corresponding to the Ornstein-Uhlenbeck process \eqref{eq:ornstein-uhlenbeck-matern-sde}. The $x$-axis shows the level of resolution, indicating the number of discretisation points and thus the number of vertices.
    }
    \label{fig:ou-kld-vs-resolution}
\end{figure}

Traditional EDA methods often suffer from model divergence as resolution increases, as discussed in Section \ref{subsec:supservised-learning-of-statistical-model}. Ideally, higher resolution brings the ensemble distribution $P$ closer to the analytical solution $P_0$, which in turn brings the statistical approximation $Q$ closer. This experiment examines whether EnIF may leverage this, and compares it with an EnKF/ES method employing the sample covariance.

To illustrate this, we select a scenario where the distributions of both the analytical solution $P_0$ and the numerical solution $P$ to Equation \eqref{eq:general-spde} are Gaussian. The $n=1000$ ensemble is generated from $P$, from which $Q$ in EnIF and ES are estimated.

We consider the 1D Matérn process described in Example \ref{ex:conditional-matern-exact}, which corresponds to a specific Ornstein-Uhlenbeck (OU) \citep{uhlenbeck1930theory} process characterised by the stochastic differential equation
\begin{align}\label{eq:ornstein-uhlenbeck-matern-sde}
    dX_t = -\kappa^{-1}X_t dt + dW_t,
\end{align}
where $W_t$ is a Wiener process, interpreted in the Itô sense \citep{oksendal2003stochastic}. The numerical solution $P$ is obtained using the Euler-Maruyama scheme \citep{platen1992numericalspde}, which has strong order of convergence 1.0 for this problem. For the OU process, the Euler scheme reduces to an AR-1 process with $\phi=1.0-\kappa^{-1}\Delta t$, leading to a multivariate Gaussian distribution for $P$, as detailed in Appendix \ref{app:auto-regressive}.

Figure \ref{fig:ou-kld-vs-resolution} displays the average KLD, $p^{-1}\KLD(P_0||Q)$, where $Q$ is represented by the Euler-Maruyama $P$ or estimated Gaussian distributions from EnIF and EnKF using the ensemble simulated under $P$.

Initially, both EnKF and EnIF benefit from the reduced divergence from $P$ to $P_0$ as resolution increases. However, statistical noise in the EnKF quickly dominates, causing the bias term $B$ in Equation \eqref{eq:bias-corrected-nll} to increase, leading $Q$ to diverge from both $P$ and $P_0$. In contrast, EnIF closely tracks the Euler scheme, eventually stabilising.

In this experiment, both EnIF and EnKF have parameter spaces that include $P$, allowing us to analyse the bias term through AIC $B(\theta)=\operatorname{len}(\theta)$. For EnIF, the number of parameters (edges) grows linearly with resolution, resulting in a linear increase in $B$, which is offset by the averaging over vertices. In comparison, EnKF effectively behaves w.r.t. a fully connected graph, where the number of edges grows quadratically with resolution, causing $B$ to increase quadratically. This leads to a linear growth in the average KLD, as shown by the green line in Figure \ref{fig:ou-kld-vs-resolution}.

\subsection{Adapting to strength of dependence}
\label{subsec:adapting-to-strength-of-dependence}
\begin{figure}[t]
    \centering
    \includegraphics[width=\textwidth]{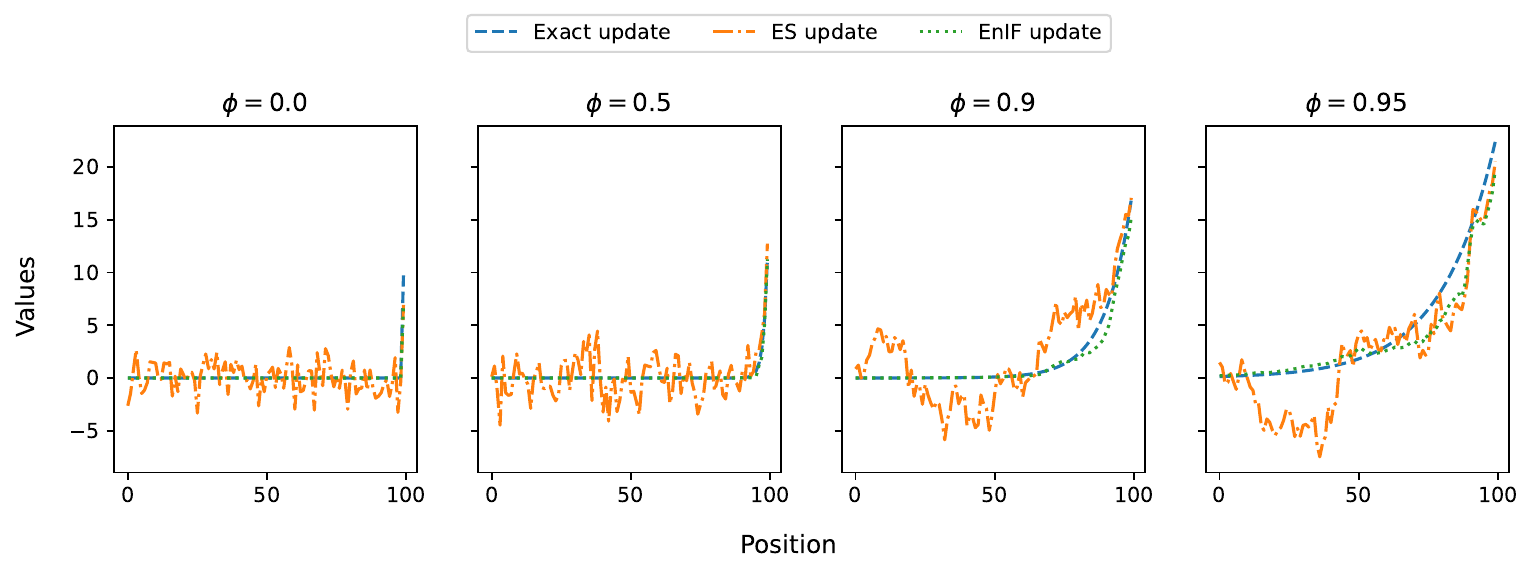}
    \caption{
    Exact, ES and EnIF updates of the first realisation in an $n=50$ ensemble of AR-1 processes with varying dependence strength, $\phi \in [0.0, 0.5, 0.9, 0.95]$. 
    }
    \label{fig:ar-1-varying-dependence-strength}
\end{figure}

The following experiment verifies that EnIF adaptively adjusts its localisation based on the strength of dependence in the ensemble, as outlined in consideration 3 of Section \ref{subsec:the-ensemble-information-filter}.

The AR-1 process (detailed in Appendix \ref{app:auto-regressive}) has Gaussian joint distribution, with its correlation function exhibiting exponential decay. By varying the dependence strength $\phi$, we can generate sample paths ranging from independent white noise ($\phi=0$) to strongly correlated data ($\phi\to 1$). We examine ensembles with $p=100$ steps and a relatively small ensemble size $n=50$ to make the statistical estimation effects more visible.

For each experiment, we assimilate an observation of the endpoint $d=20$, with a standard normal observation noise. This observation, chosen far from the prior, combined with small observation noise, increases the innovation term of the Kalman filter. This will highlight statistical inconsistencies, making them easier to observe.

In Figure \ref{fig:ar-1-varying-dependence-strength} we compare updates from the exact solution, EnIF, and EnKF/ES, across different values of  $\phi$ ($\phi \in [0.0, 0.5, 0.9, 0.95]$).

The EnKF/ES update shows significant unwanted random behaviour, especially when dependence is weak. For instance, in the left-most panel ($\phi=0$), where there is no true correlation, the exact solution only updates the last value, while the rest of the state retains the prior. In contrast, EnKF/ES updates all values noisily, introducing unnecessary variations where no correation exists.

EnIF, by contrast, demonstrates a built-in adaptive localisation due to the use of the graph $\mathcal{G}$. Regardless of the dependence strength, EnIF provides smoother and more reliable updates, closely tracking the exact solution. While not entirely free from statistical error, EnIF's updates are consistently more accurate, especially at points of weak dependence and smaller updates.

In all cases, EnIF's deviations from the exact update are minimal, demonstrating its ability to automatically adapt to varying strengths of dependence without requiring manual tuning, unlike traditional localisation methods.

\subsection{Implicit optimal localisation}
\label{subsec:implicit-optimal-localisation}
\begin{figure}[t]
    \centering
    \includegraphics[width=0.6\textwidth]{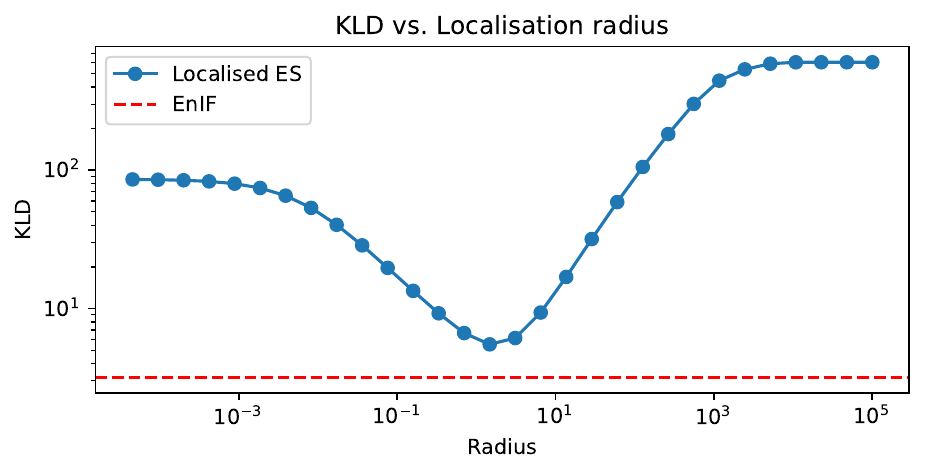}
    \caption{
    KLD values \eqref{eq:kullback-leibler-gaussian} vs. localisation radius of ES/EnKF. Both the $x$- and $y$-axes are on a log-scale.
    }
    \label{fig:matern-optimal-localisation}
\end{figure}

In the previous subsection, we demonstrated that ES/EnKF requires something to mitigate spurious correlations. In practice, most ES and EnKF-type methods employ some form of localisation, which, while ad-hoc and lacking theoretical justification, greatly improves their performance. To provide a meaningful comparison, it is necessary to compare EnIF with a localised ES/EnKF.

We consider the same Matérn model and experimental setup as in Section \ref{subsec:convergence-to-the-true-solution}, focusing on the highest resolution example with $p=1000$ and a large ensemble size of $n=1000$ to minimise the effect of statistical errors. With a fixed $\bm{H}$, we use covariance localisation \citep{houtekamer2001sequential}, varying the radius of influence $c$ from small to large. We then compare the localised ES/EnKF statistical model $Q$ to the true solution $P_0$ using the Gaussian KLD \eqref{eq:kullback-leibler-gaussian} across different values of $c$, and also compute the KLD to $P_0$ for the EnIF statistical model $Q$.

Figure \ref{fig:matern-optimal-localisation} presents the results. At very small values of $c$, the localisation effect is negligible, and $Q$ behaves like the vanilla ES/EnKF. As $c$ increases, the localisation effect improves the KLD, reaching an optimal value for the localised ES (blue line). However, further increases in $c$ weaken the udpates too much, eventually reducing the covariance matrix to a diagonal structure with only sample variances. Across all values of $c$, EnIF achieves a lower KLD.

EnIF inherently encodes how information should propagate through the system. In models governed by local operators, such as those described by Equation \eqref{eq:general-spde}, information cannot "teleport" across distant state variables -- it must propagate through the system. EnIF captures this through its Markov properties, which restrict the degrees of freedom during estimation, thereby minimising $B$ and focusing on relevant regions in the space of covariance matrices. In contrast, the ES/EnKF methods allow information to propagate unrealistically, and while localisation dampens this effect, it remains suboptimal because the structure should ideally be imposed before any statistical estimation occurs.

\section{Applications to smoothing, filtering, and parameter estimation}
\label{sec:applications}

This section illustrates how EnIF adapts to different structural regimes arising in practice. In filtering problems, conditional independence must typically be introduced as a parsimonious approximation, with optimality depending on $n$. In contrast, smoothing problems often admit exact Markov structure induced by local dynamics. Finally, in high-dimensional parameter estimation, sparsity must be learned under severe statistical constraints. Together, these examples demonstrate how EnIF accommodates both model-derived and data-driven structure in a unified framework.


\subsection{Filtering: Lorenz-96 dynamics with Runge-Kutta-4 integration}
\label{subsec:filtering-lorenz-96}
\begin{figure}[t]
    \centering
    \includegraphics[width=0.9\textwidth]{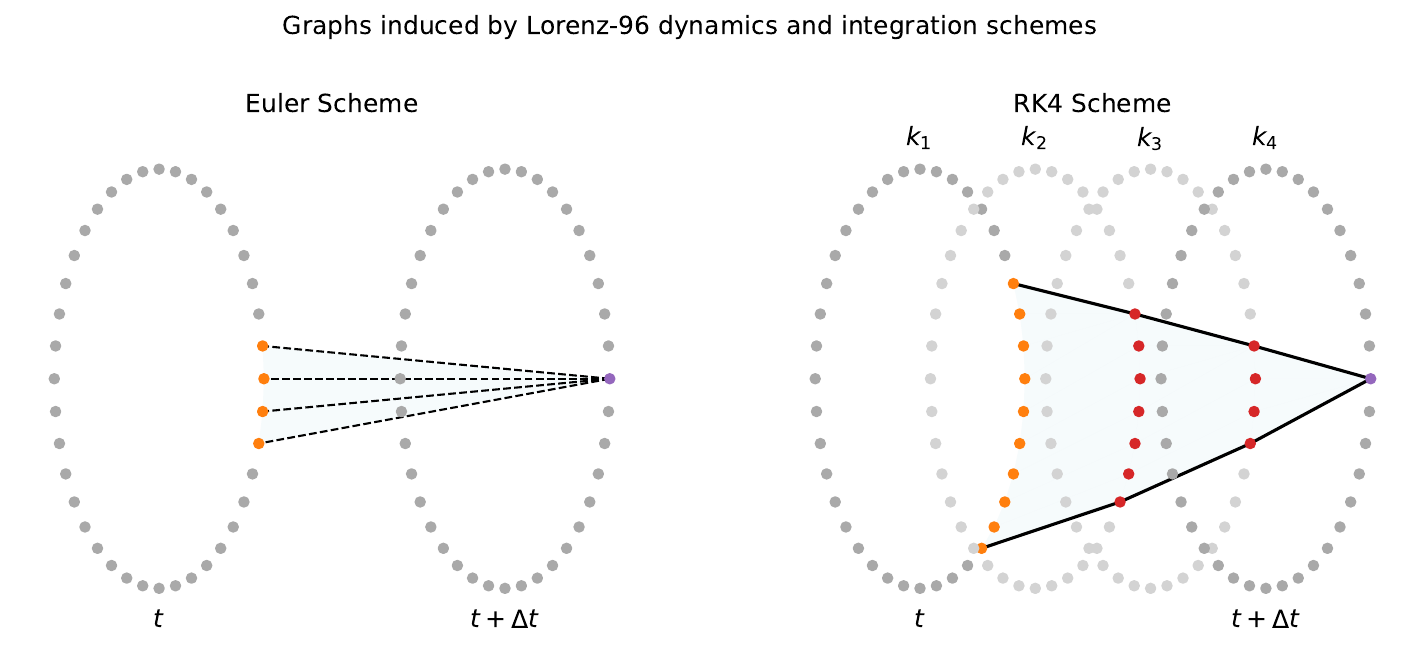}
    \caption{
    Conditional independence graphs induced by integrating the Lorenz-96 dynamics (with 40 states) in Equation \eqref{eq:model-lorenz96} using the Euler scheme (left) and the RK4 scheme (right).
    }
    \label{fig:lorenz-96-smoothing-graph}
\end{figure}
\begin{figure}[t]
    \centering
    \includegraphics[width=\textwidth]{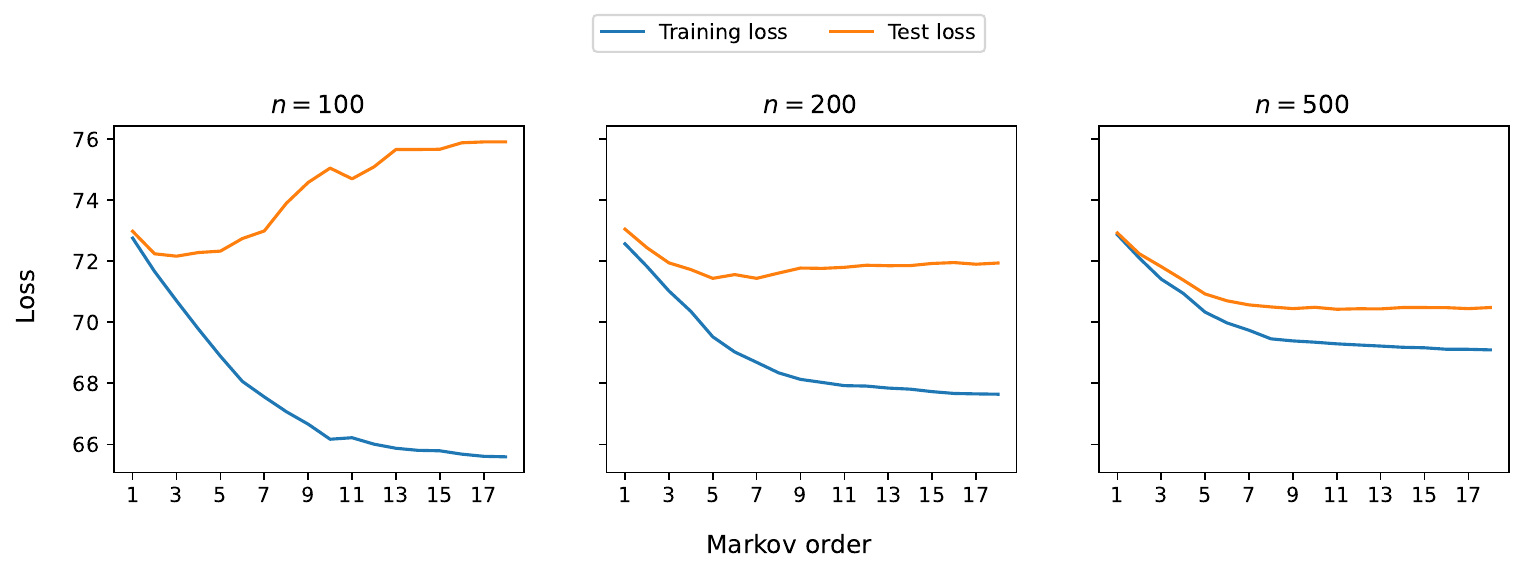}
    \caption{
Negative log-likelihood loss versus Markov order for the Lorenz-96 filtering problem. Loss is computed for training and test sets of size 
$n\in[100, 200, 500]$.
    }
    \label{fig:lorenz-96-markov-order}
\end{figure}

The Lorenz-96 model \citep{lorenz1996predictability, lorenz1998optimal} is a deterministic dynamical system describing the \( m \)-dimensional state variable \( \bm{x}(t) \), evolving as:

\begin{align}\label{eq:model-lorenz96}
    \frac{d}{dt}x_j = (x_{j+1} - x_{j-2})x_{j-1} - x_j + F,
\end{align}

where \( x_{-1} = x_{m-1} \), \( x_0 = x_m \), \( x_{m+1} = x_1 \), and \( m \geq 4 \) (here \( m = 40 \)). The constant forcing term \( F = 8 \) generates chaotic dynamics.

The Lorenz-96 model is a standard benchmark in EDA literature. We follow a similar setup to \citet{ramgraber2023ensemble2}, initializing \( \bm{x}_0 \sim 0.01\bm{z} \), where \( \bm{z} \) is standard Gaussian. Using the Runge-Kutta 4 (RK4) scheme, the system is integrated up to \( t = 4.0 \), generating training and test sets of size \( n \in [100, 200, 500] \).

The conditional independence graphs induced by integrating Equation \eqref{eq:model-lorenz96} using the Euler and RK4 schemes differ. For a visual reference, see Figure \ref{fig:lorenz-96-smoothing-graph}. The Euler scheme connects \( \bm{x}_j(t+\Delta t) \) with \( \bm{x}_{j-2}(t) \), \( \bm{x}_{j-1}(t) \), \( \bm{x}_{j}(t) \), and \( \bm{x}_{j+1}(t) \), while the RK4 scheme incorporates longer-range dependencies, connecting \( j-6 \) through \( j+3 \). For smoothing, estimating \( \bm{\Lambda} \) using the RK4 graph ensures all relevant connections are included. However, at smaller sample sizes, restricting \( \bm{\Lambda} \) to the simpler Euler graph can reduce statistical noise.

For filtering, no nodes are truly conditionally independent. Following Section~\ref{subsec:implications-for-eda}, we construct a parsimonious circular graph where node \( j \) connects only to \( j-1 \) and \( j+1 \). The Markov order is then incrementally increased by including neighbors' neighbors, and the loss (Equation \eqref{eq:negative-log-likelihood-empirical-distribution-relative-kld}) is evaluated on training and test sets.
The evaluations are presented in Figure \ref{fig:lorenz-96-markov-order}.

The results align with the theoretical predictions of Section~\ref{subsec:implications-for-eda}. The optimal Markov order, minimising test loss, increases with sample size, reaching approximately 2, 5, and 7 for \( n = 100 \), \( n = 200 \), and \( n = 500 \), respectively. At infinite ensemble size, a fully connected graph (as in the EnKF) would be correct, but finite ensembles require avoiding overfitting. By leveraging the induced local kernel effect in filtering, EnIF effectively balances parsimony and model accuracy. 

This illustrates the filtering regime, where conditional independence must be introduced as a parsimonious approximation, with the optimal level of sparsity depending on the available sample size.

\subsection{Smoothing: Exact Markov structure from local dynamics}

In contrast to the filtering setting, where conditional independence is generally approximate, smoothing problems often admit exact Markov structure induced by the underlying dynamics. This occurs when the governing equations are local, as discussed in Section~\ref{sec:markov-structure-from-local-operators}.

To illustrate this, consider the 2D stochastic heat equation
\begin{align}
    \frac{\partial u}{\partial t} - \alpha \laplacian u = \sigma \mathcal{W},
\end{align}
where \(u_t(\bm{x})\) is the temperature field and \(\mathcal{W}\) denotes white noise. Discretising the spatial domain using a triangular mesh yields basis functions with local support, such that interactions occur only between neighbouring nodes.

For example, the local mass matrix over a triangular element \(T_e\) is given by
\begin{align}
    M^{(e)} = \frac{\text{Area}(T_e)}{12}
    \begin{pmatrix}
        2 & 1 & 1 \\
        1 & 2 & 1 \\
        1 & 1 & 2
    \end{pmatrix},
\end{align}
which couples only the three nodes of the element. Using the common diagonal approximation
\begin{align}
    \tilde{M}^{(e)} = \frac{\text{Area}(T_e)}{3}\bm{I}_3,
\end{align}
preserves this locality while simplifying the resulting precision.

A similar locality property holds for the stiffness matrix, whose entries involve inner products of gradients of the same locally supported basis functions and therefore also couple only neighbouring nodes. Consequently, both the mass and stiffness matrices are sparse. A full derivation is provided in Appendix~\ref{app:subsec-stochastic-heat-equation-FEM-spde}.

As a result, the precision matrix in Equation~\eqref{eq:smoothing-precision-fem-euler} is sparse, reflecting exact conditional independence relationships between neighbouring states. This sparsity is inherited directly from the local dynamics and discretisation, rather than imposed as an approximation.

This highlights an important distinction: while filtering requires learning a parsimonious approximation to the dependence structure, smoothing may inherit exact sparsity from the model itself. In such cases, EnIF aligns exactly with the underlying probabilistic structure.

Furthermore, as in the filtering setting, the precision matrix need not be derived explicitly from the governing equations. It may instead be estimated from the ensemble with respect to a graph encoding only the locality, allowing EnIF to recover or approximate the underlying structure directly from data when an explicit model is unavailable.
This represents the complementary regime, where sparsity is not learned or imposed, but inherited exactly from the underlying model.

\subsection{Static parameters: High-dimensional non-Markov GRFs with unknown observation operator}
\label{subsec:static-parameters-large-2d-matern-grf}

\begin{figure}[t]
    \centering
    \includegraphics[width=0.9\textwidth]{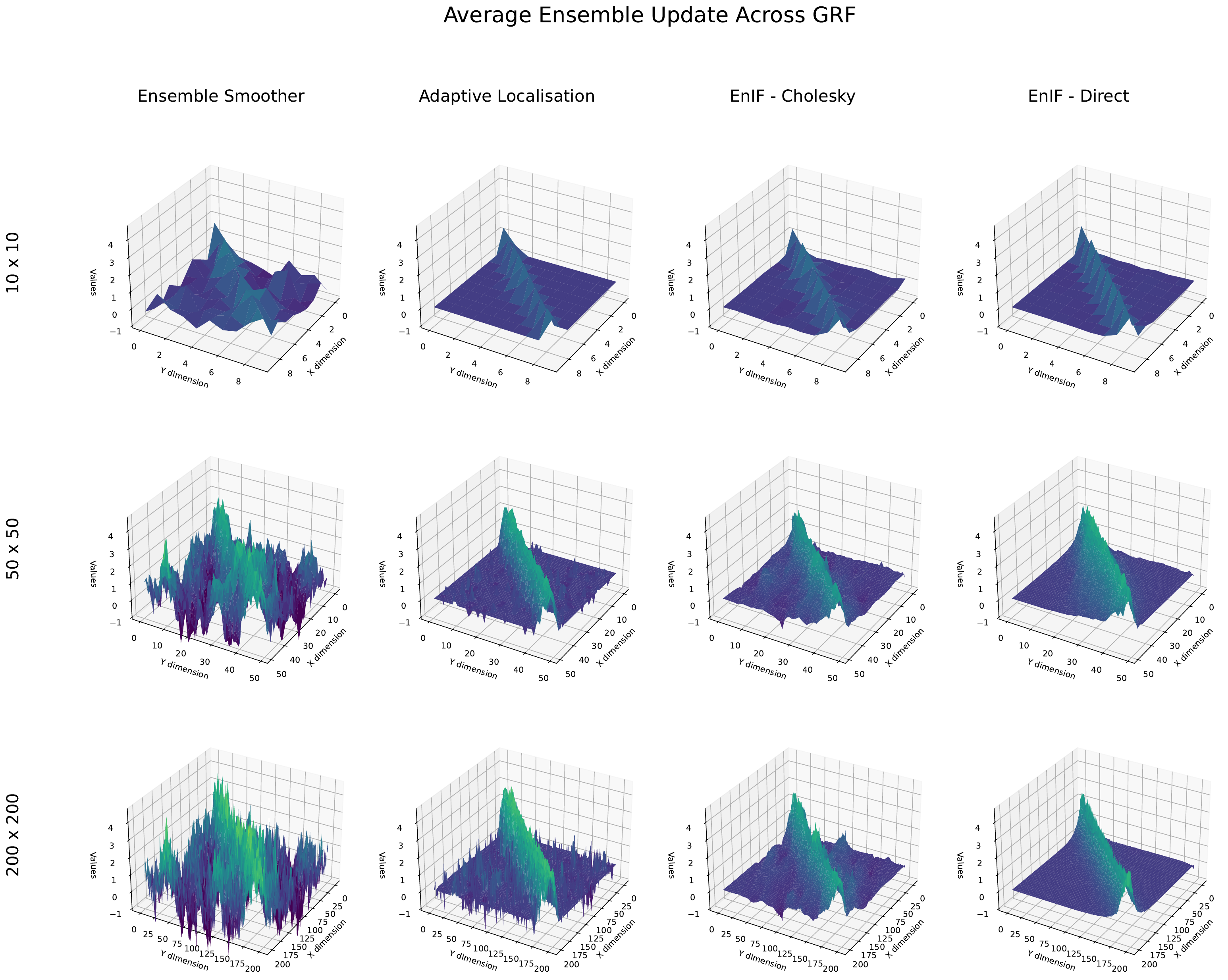}
    \caption{
   The average ($n=100$) ensemble update across the GRFs at varying discretisation, using different update algorithms. The GRFs are sampled using an anisotropic exponential covariance function. This is a special case of the Matérn covariance, which in two dimensions does not yield a Markov property. The response is sampled as $y = u + z$ for $z \sim \mathcal{N}(0, 0.1^2)$, i.e. direct but noisy observations along the diagonal elements in $\bm{u}$. The true $\bm{H}$ is sparse but unknown, and must be estimated from the ensemble.
    }
    \label{fig:large-anisotropic-exponential-grf}
\end{figure}

We consider a high-dimensional parameter estimation problem motivated by history matching in reservoir engineering, where the goal is to update spatially distributed parameters representing geological uncertainty. These parameters are commonly modelled using Gaussian random fields (GRFs), which exhibit strong spatial locality but do not, in general, admit exact Markov structure.

This setting combines several key challenges:

\paragraph{Non-Markov dependence.}
The GRF exhibits strong spatial locality but does not admit an exact Markov structure in two dimensions. Consequently, no exact conditional independence graph exists, and any sparse representation must be imposed as a parsimonious approximation.

\paragraph{High dimensionality $p$.}
The parameter dimension \(p\) grows rapidly with spatial resolution, leading to a \(p \gg n\) regime where classical covariance estimation becomes severely ill-posed.

\paragraph{Limited ensemble size $n$.}
The ensemble size \(n\) is constrained by the computational cost of forward simulations, further exacerbating statistical instability and overfitting.

\paragraph{Unknown observation operator.}
The observation operator \( \bm{h} \) is not known explicitly and must be estimated from the ensemble. This implies that its linearisation \( \bm{H} \) must be learned as part of the update, turning the assimilation step into a coupled problem of high-dimensional precision estimation and sparse regression.

To reflect these challenges, we construct a synthetic example that captures all four aspects simultaneously. We sample \(n = 100\) realisations from a 2D GRF with an anisotropic exponential covariance function. While this covariance exhibits strong locality, it does not correspond to a Markov random field in two dimensions. We therefore impose a parsimonious graph structure by connecting only neighbouring nodes, enforcing a local approximation to the true dependence structure.

The GRFs are sampled at three discretisation levels: \(10 \times 10\), \(50 \times 50\), and \(200 \times 200\), corresponding to 100, 2,500, and 40,000 parameters, respectively. Observations are generated as
\[
y = u + z, \quad z \sim \mathcal{N}(0, 0.1^2),
\]
corresponding to direct but noisy observations along the diagonal of \( \bm{u} \). While the true population \( \bm{H} \) is sparse, this structure is not provided to the algorithms and must instead be estimated from the ensemble.

This setup therefore tests whether EnIF can simultaneously (i) recover a parsimonious approximation to a non-Markov dependence structure, and (ii) estimate a high-dimensional observation operator under limited data.

We evaluate updates from four methods:
\begin{enumerate}
    \item Ensemble Kalman Smoother (ES).
    
    \item ES with adaptive localisation using a threshold $3\sqrt{n}^{-1}$.
    
    \item EnIF with precision estimated via a permutation-optimised affine KR-map 
    (approximately a Cholesky factor; see Section~\ref{subsec:encoding-conditional-independence-from-G-inthe-KR-map-S}).
    
    \item EnIF with a directly estimated precision matrix \citep{lunde2022graphspme}.
\end{enumerate}

The average updates,
\[
n^{-1} \sum_i (\bm{u}_{t|t}^{(i)} - \bm{u}_{t|t-1}^{(i)}),
\]
are shown in Figure~\ref{fig:large-anisotropic-exponential-grf}.

As the resolution increases, the statistical challenges become more pronounced. At \(200 \times 200\) (40,000 parameters), the ES update fails to distinguish signal from noise, resulting in highly irregular updates. Adaptive localisation mitigates this issue but lacks an intrinsic preference for locality, leading to spurious long-range effects and unstable behaviour.

In contrast, EnIF produces stable and spatially coherent updates even at high resolution. By enforcing sparsity in the precision matrix with respect to a local graph, EnIF effectively captures the dominant dependence structure despite the absence of exact conditional independence. At the same time, sparse estimation of \( \bm{H} \) ensures that the observation operator is learned in a statistically efficient manner.

The Cholesky-based variant introduces additional degrees of freedom due to fill-in, resulting in slightly irregular updates, while the direct precision estimation approach avoids this effect but may require higher Markov order and careful handling to ensure positive definiteness.

These results highlight the importance of jointly regularising both the precision matrix and the observation operator in high-dimensional settings. In this regime, where neither exact model structure nor sufficient data is available, EnIF provides a principled framework for learning parsimonious approximations directly from the ensemble.

\section{Discussion}
\label{sec:discussion}

This paper addresses fundamental statistical limitations of ensemble-based data assimilation (EDA), in particular the reliance on localisation to mitigate spurious correlations and ensemble collapse in high-dimensional, low-sample regimes. By combining ideas from triangular measure transport, supervised learning, and SPDE-based modelling, we show that locality in the underlying system provides a principled basis for regularisation through sparse precision structure.

The proposed Ensemble Information Filter (EnIF) encodes this structure directly, replacing covariance-based representations with sparse precision parametrisations. This yields a method that is both statistically and computationally efficient, and which avoids the need for heuristic localisation. Rather than suppressing spurious correlations after estimation, EnIF incorporates structural assumptions prior to estimation, leading to improved stability and reduced bias.

A key insight of this work is that approximate Markov structure, even when not exact, provides an effective and often optimal regularisation in the Kullback–Leibler sense. This was demonstrated on filtering problems, where EnIF adapts to the strength and range of dependencies in the data. In particular, the history matching parameter estimation example highlight how sparse representations remain effective even when the true system is not strictly Markovian, provided there is a preference for locality.

In addition to precision estimation, the methodology also addresses the learning of the observation operator through sparse regression. This is particularly important in regimes where the observation mapping is unknown or must be linearised, and where $p \gg n$ makes standard estimation ill-posed. By enforcing sparsity in $\bm{H}$, the framework maintains consistency with the assumed conditional independence structure and ensures that the posterior precision remains sparse, which is critical for both statistical and computational scalability.

From a practical perspective, the methodology requires only a specification of the state and observation models together with a graph encoding conditional independence. This separates modelling assumptions from estimation, and enables scalable inference in regimes where $p \gg n$.

Several directions for future work follow naturally. Extending the framework to non-linear KR maps would allow fully non-Gaussian and non-linear inference while preserving sparsity and scalability. Furthermore, improvements in the underlying learning procedures -- particularly sparse regression for estimating $\bm{H}$ and precision estimation under extreme dimensionality -- may be required in applications with low signal-to-noise ratios. In such settings, leveraging additional structure, such as the computational graph of $\bm{h}$ or spatially informed regression, may be essential.

\appendix
\section{Information filter update equivalent to Kalman update}
\label{app:enif-enkf-equivalence}
The Ensemble Kalman filter update equation is 
\begin{align}
    \bm{u}_{t|t}^{(i)} = \bm{u}_{t|t-1}^{(i)} + \bm{K}(\bm{d}_t - \bm{d}_{t|t-1}^{(i)}) 
    = \bm{u}_{t|t-1}^{(i)} + \bm{K}(\bm{d}_t - (\bm{y}_{t|t-1}^{(i)}+\bm{\epsilon}^{(i)}))
\end{align}
where the Kalman gain is
\begin{align}
    \bm{K} = \bm{\Sigma_{uy}}\bm{\Sigma_{d}}^{-1} = \bm{\Sigma_{t|t-1}}\bm{H}^\top\left(\bm{\Sigma_{y}}+\bm{\Sigma_{\epsilon}}\right)^{-1}.
\end{align}
In the following we show how to produce the same update but through the Information filter update equation.
The Ensemble Information filter update is
\begin{itemize}
    \item Rescale each realisation into the "canonical" space
    \begin{align}
        \bm{\eta}_{t|t-1}^{(i)} = \Lambda_{t|t-1} \bm{u}_{t|t-1}^{(i)}.
    \end{align}
    \item Update each rescaled realisation using the IF equations, along with an update of the precision
    \begin{align}
        \bm{\eta}_{t|t}^{(i)} &= \bm{\eta}_{t|t-1}^{(i)} + \bm{H}^\top \bm{\Lambda}_{\bm{r}_t}\tilde{\bm{d}}_t^{(i)}  \\
        \bm{\Lambda}_{t|t} &= \bm{\Lambda}_{t|t-1} + \bm{H}^\top \bm{\Lambda}_{\bm{r}_t} \bm{H} .
    \end{align}
    \item Bring each rescaled realisation back to its original space
    \begin{align}
        \bm{u}_{t|t}^{(i)} = \Lambda_{t|t}^{-1} \bm{\eta}_{t|t}^{(i)}.
    \end{align}
\end{itemize}

To see that the Kalman update is the same as the Information filter update, note firstly that
\begin{align}
    \bm{\Lambda}_{t|t}^{-1}\bm{\eta}_{t|t-1}^{(i)}
    &= \bm{\Lambda}_{t|t}^{-1} \bm{\Lambda}_{t|t-1} \bm{u}_{t|t-1}^{(i)} \notag \\
    &= \left(\bm{\Sigma}_{t|t-1} - \bm{K} \bm{H}\bm{\Sigma}_{t|t-1}\right) \bm{\Sigma}_{t|t-1}^{-1} \bm{u}_{t|t-1}^{(i)} \notag \\
    &= \bm{u}_{t|t-1}^{(i)} - \bm{K}\hat{\bm{y}}_{t|t-1}^{(i)},
\end{align}
where $\hat{\bm{y}}_{t|t-1}^{(i)} = \bm{H}\bm{u}_{t|t-1}^{(i)}$.
Secondly, note that 
$\bm{K} = \bm{\Sigma}_{t|t}\bm{H}^\top \bm{\Lambda}_{\bm{r}_t}$
from an application of Woodbury matrix inversion formula (see e.g. \citet{moore1979optimal}).
Therefore
\begin{align}
    \bm{\Lambda}_{t|t}^{-1}\bm{H}^\top \bm{\Lambda}_{\bm{r}_t}\tilde{\bm{d}}_t^{(i)} = \bm{K}\tilde{\bm{d}}_t^{(i)}.
\end{align}
Define the noisy residual as 
$\bm{r}^{(i)}=\bm{h}(\bm{u}_{t|t-1}^{(i)})-\bm{H}\bm{u}_{t|t-1}^{(i)}+\bm{\epsilon}$,
where $\bm{\epsilon}$ is zero-mean Gaussian noise
$\bm{\epsilon}_i$
sampled according to a precision specifying observation uncertainty $\bm{\Lambda}_{\bm{\epsilon}_t}$.
Thus $\bm{\Lambda}_{\bm{r}_t}$ specifies all uncertainty in a prior simulation of the uncertain observation 
$\bm{d}_{t|t-1}^{(i)}=\bm{h}(\bm{u}_{t|t-1}^{(i)})+\bm{\epsilon}^{(i)}$
not explained by the regression $\bm{H}\bm{u}_{t|t-1}^{(i)}$.
Finally, let 
$\tilde{\bm{d}}_t^{(i)}=\bm{d} - \bm{r}^{(i)}$.
Then we see that
\begin{align}
    \bm{\Lambda}_{t|t}^{-1} \bm{\eta}_{t|t}^{(i)}
    &= \bm{\Lambda}_{t|t}^{-1}\left( \bm{\eta}_{t|t-1}^{(i)} + \bm{H}^\top \bm{\Lambda}_{\bm{r}_t}\tilde{\bm{d}}_t^{(i)} \right) \notag \\
    &= \bm{u}_{t|t-1}^{(i)} - \bm{K}\hat{\bm{y}}_{t|t-1}^{(i)} + \bm{K}\tilde{\bm{d}}_t^{(i)} \notag \\
    &= \bm{u}_{t|t-1}^{(i)} + \bm{K}\left( \bm{d} - (\bm{h}(\bm{u}_{t|t-1}^{(i)})+\bm{\epsilon}^{(i)}) \right) \notag \\
    &= \bm{u}_{t|t}^{(i)}.
\end{align}

\section{The AR-$1$ process}
\label{app:auto-regressive}
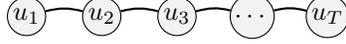
\begin{figure}
    \centering
    \begin{tikzpicture}
    \tikzstyle{vertex}=[circle,draw=black,fill=black!5,minimum
    size=12pt,inner sep=1pt]
    \node[vertex](1) at (0,0) {$u_1$};
    \node[vertex](2) [right of=1] {$u_2$};
    \node[vertex](3) [right of=2] {$u_3$};
    \node[vertex](D) [right of=3] {$\cdots$};
    \node[vertex](T) [right of=D] {$u_T$};
    \path[draw,thick,-] (1) to[out=15,in=165] (2);
    \path[draw,thick,-] (2) to[out=15,in=165] (3);
    \path[draw,thick,-] (3) to[out=15,in=165] (D);
    \path[draw,thick,-] (D) to[out=15,in=165] (T);
    \end{tikzpicture}
    \caption{The graph corresponding to an AR-1 process}
    \label{fig:auto-regressive-1}
\end{figure}

The AR-1 process is defined by
\begin{align}\label{eq:auto-regressive-1}
    u_t=\phi u_{t-1}+\epsilon_t,~ 
    u_1\sim\mathcal{N}\left(0,\frac{1}{1-\phi^2}\right),~
    \epsilon_t\sim \mathcal{N}(0,1).
\end{align}
Here, some $u_t$ is "generated" only by $u_{t-1}$, 
implying a first order connection formalised as edges only between sequential elements 
as shown in Figure \ref{fig:auto-regressive-1}.
Correspondingly, the precision matrix of $\bm{u}=[u_1,\ldots,u_T]^\top$ is tridiagonal and sparse,
in contrast to the dense covariance
\begin{align}\label{eq:auto-regressive-1-cov-prec}
\bm{\Sigma}_{\bm{u}} &=
\begin{bmatrix}
B(1,1) & \cdots & B(1,T)\\
\vdots & \ddots & \vdots \\
B(T,1) & \cdots & B(T,T) 
\end{bmatrix},~
B(i,j)=\frac{\phi^{|i-j|}}{1-\phi^2},~~~
\Lambda_{\bm{u}} =
\begin{bmatrix}
1 & -\phi \\
-\phi & 1+\phi^2 & -\phi \\
& \ddots & \ddots & \ddots \\
& & -\phi & 1+\phi^2 & -\phi \\
& & & -\phi & 1
\end{bmatrix}
\end{align}
formed by considering that $u_t$ depends upon $u_{t-1}$ which depends upon $u_{t-2}$ and so on.
\section{The Matérn-1/Ornstein-Uhlenbeck analytical and approximate solutions}
\label{app:ornstein-uhlenbeck-solution}

The OU SDE in Equation \eqref{eq:ornstein-uhlenbeck-matern-sde} correspond to the specific 1D Matérn SPDE in Equation \eqref{eq:matern-spde-1d} with $\kappa=1.0$. Therefore, the stationary covariance between time-points $\Delta t$ is $\frac{1}{2}e^{-\Delta t}$.

When integrated approximately using the Euler-Maruyama scheme, the process follows the AR-1 dynamics in Appendix \ref{app:auto-regressive}, with $\phi=1-\Delta t$.
\section{Applying fill-in reduction algorithms on $C^\top C=\Lambda$}
\label{app:in-fill-reduction-on-CTC}

We have that $\bm{C}$ is lower triangular and that $\bm{\Lambda}_{\bm{u}} = \bm{C}^\top \bm{C}$.
In-fill reduction algorithms like the AMD ordering \citep{amestoy2004algorithm}, Reverse-Cuthill McKee ordering \citep{cuthill1969reducing}, or METIS \citep{karypis1997metis}
are typically implemented for the ordinary Cholesky decomposition $\bm{\Lambda} = \bm{L} \bm{L}^\top$
where $\bm{L}$ also is lower triangular.
We seek to exploit these algorithms for finding an ordering $\pi$ of $\bm{u}$ so that $\bm{C}(\pi)$
has as little in-fill as possible compared to that of $\bm{\Lambda}_{\bm{u}}$.

First, note that the Cholesky decomposition is unique because $\bm{\Lambda}_{\bm{u}}$ is positive definite.
Let $\bm{P}_r$ be the symmetric permutation matrix reversing the order, $\pi_r=(p,\ldots,1)$,
then $\bm{P}_r=\bm{P}_r^\top=\bm{P}_r^{-1}$ and $\bm{P}_r\bm{P}_r=\bm{I}$.
We have that
\begin{align}\label{eq:reverse-perm-L-to-C}
    \bm{P}_r \bm{\Lambda}_{\bm{u}} \bm{P}_r
    = 
    (\bm{P}_r \bm{C}^\top \bm{P}_r)
    (\bm{P}_r \bm{C}^\top \bm{P}_r)^\top.
\end{align}
Thus from uniqueness $\bm{P}_r \bm{C}^\top \bm{P}_r=\tilde{\bm{L}}$ must be a lower triangular Cholesky factor of the reversed precision 
$\hat{\bm{\Lambda}}_{\bm{u}} = \bm{P}_r \bm{\Lambda}_{\bm{u}} \bm{P}_r$, since this too is SPD.
Note that $\tilde{\bm{L}}$ will have the same amount of in-fill as $\bm{C}$, as the reverse permutation changes nothing here.
Therefore, we may first seek an additional optimised permutation, say $\bm{P}_*$, so that the Cholesky factor, say $\bm{L}_*$, has as little extra non-zeroes compared to $\bm{\Lambda}_{\bm{u}}$ as possible.
As such, seek an optimized ordering $\bm{P}_*$ and corresponding sparsity structure of a Cholesky factor $\bm{L}_*$ as
\begin{align}
    \bm{L}_*\bm{L}_*^\top = \bm{P}_*^\top \bm{\Lambda}_{\bm{u}} \bm{P}_*.
\end{align}
We may then find the sparsity structure of $\bm{C}(\pi^*)$ from using \eqref{eq:reverse-perm-L-to-C}:
\begin{align}
    \bm{C}(\pi^*) = \left(\bm{P}_r\bm{L}_*\bm{P}_r\right)^\top.
\end{align}
The elements of $\bm{C}(\pi^*)$ may then be learned from data optimising \eqref{eq:triangular-transport-objective} only w.r.t. the non-zero elements of $\bm{C}(\pi^*)$.
To "unwrap" and retrieve the original precision, we may calculate it as
\begin{align}
    \bm{\Lambda}_{\bm{u}} = 
    \bm{P}_*
    \bm{P}_r
    \bm{C}(\pi^*)^\top \bm{C}(\pi^*)
    \bm{P}_r
    \bm{P}_*^\top.
\end{align}

\section{Fast regularised sparse regression through boosting with linear base learners}
\label{app:fast-regularised-sparse-regression-through-boosting}

\begin{figure}
    \centering
    \includegraphics[width=0.5\textwidth,keepaspectratio]{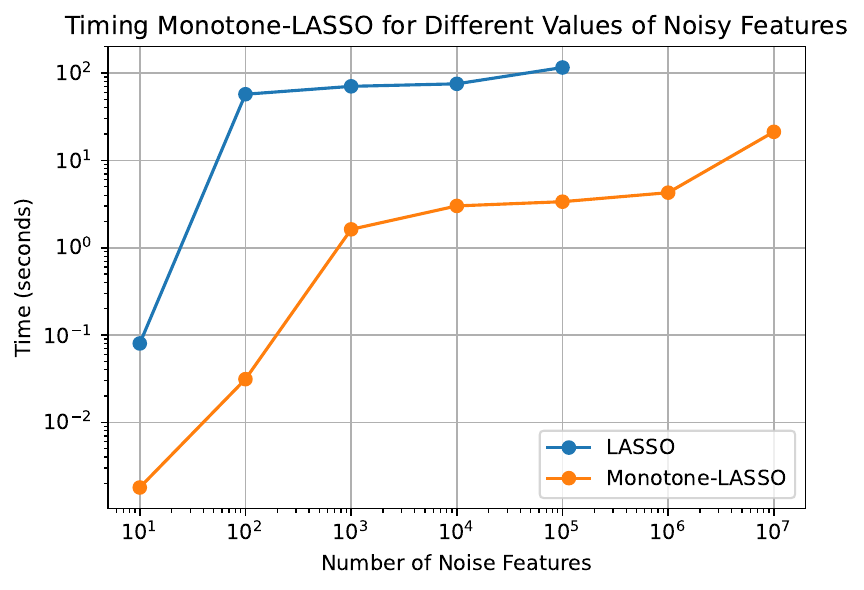}
    \caption{
Timing LASSO from the python package sklearn and the Monotone LASSO in Algorithm \ref{alg:monotone-lasso-with-early-stopping}, versus an increasing number of (noisy) regression features.
    }
    \label{fig:lasso-monotone-timing}
\end{figure}

\citet{raanes2023review} connects the "average sensitivity" of the estimated $\bm{H}$ in EnKF-type methods with the expected gradient, $\E{\nabla \bm{h}(\bm{u})}$, when $n\to\infty$. Thus the population estimate is indeed the expected gradient. We now describe the underlying elements to Algorithm \ref{alg:monotone-lasso-with-early-stopping}, which may be used when $\bm{H}$ is unknown and the dimension of $\bm{u}$ is large. It was used to produce the results in Section \ref{subsec:static-parameters-large-2d-matern-grf}.

We estimate $\bm{H}$ row-by-row. Thus, it is sufficient to consider only a vector of responses $\bm{y}$, and regression coefficients $\bm{\beta}$ eventually inserted as the row. We thus assume $\bm{H}$ is $1\times p$ in the following.

First, the data-matrix $\bm{U}=[\bm{u}^{(1)} \ldots \bm{u}^{(n)}]^\top$ is standardised. This is to ensure features are comparable at the same scale, and weighted equally by regularisation, see e.g. \citet{hastie2009elements} and the LASSO. The response vector is also standardised, but for numerical reasons. This means that estimated coefficients $\hat{\bm{\beta}}_j$ needs a rescaling before insertion into $\bm{H}$, i.e. $\hat{\bm{H}}_{1,j}=\hat{\sigma}_{y}\hat{\sigma}_{u_j}^{-1}\hat{\bm{\beta}}_j$.

To retain sparsity of the posterior precision, $\bm{\Lambda}_{t|t}$, it is important to seek a sparse estimate $\hat{\bm{H}}$. However, for statistical estimation reasons we could have chosen e.g. Ridge regression \citep{hoerl1970ridge} yielding a dense estimate. Note however also the "always bet on sparsity" principle \citep{hastie2009elements}, and the added explainability element of how certain responses effects only a few state-elements directly.

The go-to sparse linear regression algorithm is the LASSO \citep{tibshirani1996regression}. There are efficient coordinate descent algorithms for solving the optimisation problem, but unfortunately, seemingly too slow for large problems like in Section \ref{subsec:static-parameters-large-2d-matern-grf} with multiple responses. See Figure \ref{fig:lasso-monotone-timing} for some timing experiments.

Instead we implement a version of the Monotone-LASSO \citep{hastie2007forward}. This may be seen as a form of boosting \citep{friedman2001greedy, mason1999boosting} with 1D linear base learners. To make it adaptive and computationally efficient, we implement early stopping using an approximation of $n$-fold cross-validation \citep{stone1974cross, claeskens2008model}. The algorithm is provided in Algorithm \ref{alg:monotone-lasso-with-early-stopping}, and implemented in pure Python here \url{https://github.com/equinor/graphite-maps}.

Figure \ref{fig:lasso-monotone-timing} shows significant speed-ups compared to the standard LASSO. Furthermore the additional regularisation is beneficial for large scale applications: For the medium $50\times 50$ 2D GRF example in Section \ref{subsec:static-parameters-large-2d-matern-grf}, the LASSO (tuned with 10-fold cross validation) produces 1255 non-zero elements in its estimate of $\bm{H}$, while the monotone LASSO produces only 50 non-zeroes. The population quantity has 50, while a dense estimate contains 125,000 elements.

Finally, if the signal-to-noise ratio cannot be overcome by ordinary (uninformed) regression, it might be beneficial to either utilise the computational graph (see e.g. \citet{kristensen2015tmb}) or knowledge regarding positions of $\bm{u}$ and $\bm{y}$ for use in weighted boosting regression.




\begin{algorithm*}[ht!]
	\begin{tabbing}
		\hspace{2em} \= \hspace{2em} \= \hspace{2em} \= \\
		{\bfseries Input}: \\
		\> - An $n\times p$ matrix of standardised feature realisations $\bm{X}$\\
		\> - An $n$-vector of response realisations $\bm{y}$\\

		{\bfseries Do}: \\
		1. {\bfseries Initialise} $\hat{\bm{\beta}}_0 = \bm{0}$ \\
		2. {\bfseries While} $\operatorname{mse}_{cv-n}(\bm{X},\bm{y};\hat{\bm{\beta}}_k)>\operatorname{mse}_{cv-n}(\bm{X},\bm{y};\hat{\bm{\beta}}_{k+1})$: \\
		
		\> $i)$ Calculate all 1D linear regressions\\
		
		\> $ii)$ Select $\beta_j$ as the one reducing training $\operatorname{mse}$ the most \\
		
		\> $iii)$ $\hat{\bm{\beta}}_{k+1,j} = \hat{\bm{\beta}}_{k,j} + \epsilon \beta_j$ \\
		
		\hspace{0.4cm}{\bfseries end While} \\
		
		3. {\bfseries Return} $\hat{\bm{\beta}}$.
	\end{tabbing}
\vspace{0.3cm}

\textbf{Notes:} 
\begin{itemize}
    \item cv-n is computed using $\hat{\theta}_{-i} \overset{n}{\to} \hat{\theta} - n^{-1}\operatorname{IF}(y_i,\bm{x}_i)$, where the influence $\operatorname{IF}$ is found using the asymptotic properties of $\hat{\beta}_j$ as an M-estimator. See cv-n and TIC relation through $\operatorname{IF}$ in \citet{claeskens2008model}.
    \item For these 1D regressions, we construct the estimate $$\operatorname{IF}(y^{(i)},\bm{x}_j^{(i)}) = -\left(y^{(i)}-\hat{\beta}_j\bm{x}_j^{(i)}\right)\bm{x}_j^{(i)}\left[-n^{-1}\sum_i (\bm{x}_j^{(i)})^2\right]^{-1}$$.
\end{itemize}

\vspace{0.5cm}
\caption{\label{alg:monotone-lasso-with-early-stopping}Monotone LASSO with information-theoretic early stopping}
\end{algorithm*}
\section{On the multiple-data-assimilation extension}
\label{app:multiple-data-assimilation extension}

We establish that EnIF may be used within as EnIF-MDA by repeating the Gaussian-linear proof of \citet{emerick2013ensemble} using the precision. This also shows that the canonical parametrisation is "natural" for conditioning a Gaussian, as the derivations here are rather straightforward.
Appendix \ref{app:enif-enkf-equivalence} showcase the equivalence of EnKF and EnIF when the covariance/precision and $\bm{H}$ are known. This may also be considered as the infinite ensemble case. Furthermore, EnIF is different in the sense that it directly updates the precision matrix. We therefore only need to show that the EnIF-MDA updated precision is equivalent to the EnIF updated precision under linear Gaussian dynamics.

Let 
$\bm{\Lambda}_{t|t-1}$ be the ordinary prior precision.
$\bm{\Lambda}_{\bm{r}}=\bm{\Lambda}_{\bm{\epsilon}}$ is the ordinary measurement uncertainty, note that $\bm{h}(\bm{u})=\bm{H}\bm{u}$ and that we therefore have equivalence.
We then define the equivalent MDA quantities as
\begin{align}
\tilde{\bm{\Lambda}}_{\bm{\epsilon}} &=
\begin{bmatrix}
\alpha_1\bm{\Lambda}_{\bm{\epsilon}} & \bm{0} \\
\bm{0} & \alpha_2\bm{\Lambda}_{\bm{\epsilon}} & \bm{0} \\
& \ddots & \ddots & \ddots \\
& & \bm{0} & \alpha_{K-1}\bm{\Lambda}_{\bm{\epsilon}} & \bm{0} \\
& & & \bm{0} & \alpha_K\bm{\Lambda}_{\bm{\epsilon}}
\end{bmatrix}
~\text{and}~
\tilde{\bm{H}} = 
\begin{bmatrix}
    \bm{H} \\ 
    \vdots \\
    \bm{H}
\end{bmatrix}
\end{align}
where $\sum_{k=1}^K \alpha_k = 1$, and we assume $K$ iterations.

From direct calculation, we obtain that the EnIF-MDA precision is equivalent to the EnIF updated precision:
\begin{align}
    \tilde{\bm{\Lambda}}_{t|t}
    &= \bm{\Lambda}_{t|t-1} + \tilde{\bm{H}}^\top \tilde{\bm{\Lambda}}_{\bm{\epsilon}} \tilde{\bm{H}} \notag \\
    &= \bm{\Lambda}_{t|t-1} + \sum_{k=1}^K \alpha_k\bm{H}^\top \bm{\Lambda}_{\epsilon} \bm{H} \notag \\
    &= \bm{\Lambda}_{t|t-1} + \bm{H}^\top \bm{\Lambda}_{\epsilon} \bm{H} \notag \\
    &= \bm{\Lambda}_{t|t}.
\end{align}

\section{The SPDE approach}
\label{app:the-spde-approach}

FEM are applicable for a larger class of models \eqref{eq:general-spde}.
The link to Markov properties are due to \citet{lindgren2011explicit} through some explicit choices.
To illustrate the approach, we make use of a simple version of \eqref{eq:general-spde}
\begin{align}\label{eq:general-spde-first-order-in-time}
    \frac{d}{dt}u_t(\bm{x}) - \mathcal{A}u_t(\bm{x}) = \sigma \mathcal{W}.
\end{align}

The SPDE approach seeks a stochastic weak solution, so that for a suitable set of test-functions $v(\bm{x})$ we seek equality in distribution for
\begin{align}\label{eq:spde-stochcastic-weak-solution}
    \int_{\Omega} v(\bm{x})
    \left(
        \frac{d}{dt}u_t(\bm{x}) - \mathcal{A}u_t(\bm{x})
    \right) \, d\bm{x} 
    \overset{d}{=} \sigma \int_{\Omega} v(\bm{x}) \mathcal{W}(t, \bm{x}) \, d\bm{x},
\end{align}
where $\Omega$ represents the spatial domain.We note that boundary conditions are omitted, as they impose constraints localised to the boundary of \( \Omega \). Therefore they do not fundamentally affect the Markov properties of the solution within the domains interior.

The FEM solution represents $u_t(\bm{x})$ via basis functions $\phi$ and values of $u$ at the vertices of a triangular mesh (discretising the field) $\bm{u}$ so that
\begin{align}
    u_t(\bm{x}) \approx \sum_{k=1}^{p_{\bm{x}}} \mathbf{u}_k(t) \phi_k(\bm{x}),
\end{align}
where internal values are triangularized.
The SPDE approach selects $\phi$ to be piecewise linear functions so that $\phi_k=1$ at vertex $k$, but zero at all other vertices.
This choice along with e.g. the Galerkin method ($v$ are chosen from the same family as the basis functions, specifically we select $v=\phi_k$) leads to the matrices
\begin{align}\label{eq:mass-stiffness-matrices-indices}
    \bm{M}_{ij} &= \int_{\Omega} \phi_j(\bm{x}) \phi_i(\bm{x}) \, d\bm{x}, \\
    \bm{A}_{ij} &= \int_{\Omega} \phi_j(\bm{x}) \mathcal{A} \phi_i(\bm{x}) \, d\bm{x}.
\end{align}
The matrix $\bm{M}$ is band-sparse due the choice of basis $\phi$, while the same applies to $\bm{A}$ if in addition $\mathcal{A}$ is a local operator.
Taking the r.h.s. of Equation \eqref{eq:spde-stochcastic-weak-solution} as the Ito integral, the solution is mean-zero Gaussian. The matrix $\bm{M}$ also appears, as 
\begin{align}
    \operatorname{Cov}\left(\sigma \int_{\Omega} \phi_i(\mathbf{x}) \mathcal{W}(\mathbf{x}) \, d\mathbf{x}, \sigma \int_{\Omega} \phi_j(\mathbf{x}) \mathcal{W}(\mathbf{x}) \, d\mathbf{x}\right) = \sigma^2 \int_{\Omega} \phi_i(\mathbf{x}) \phi_j(\mathbf{x}) \, d\mathbf{x} = \sigma^2 \bm{M}_{ij}.
\end{align}
All spatial white-noise is integrated out, so we are left only with white-in-time noise. 
Thus, the weak solution obtains the sparse SDE system
\begin{align}
    \bm{M} \frac{d\mathbf{u}}{dt} - \bm{A}\mathbf{u} = \sigma \bm{M}^{\frac{1}{2}} d\mathbf{Z},
\end{align}
where $\bm{Z}\sim \mathcal{N}(0,\bm{I})$ is due to independent white noise in time.
A simple Euler-Maruyama method \citep{platen1992numericalspde} then achieves
\begin{align}
    \bm{M} \frac{\mathbf{u}_{t+1} - \mathbf{u}_t}{\Delta t} - \bm{A}\mathbf{u}_t = \sigma \Delta t \bm{M}^{\frac{1}{2}} \mathbf{Z}.
\end{align}
Solving, we then see an auto-regressive relationship in time
\begin{align}
    \mathbf{u}_{t+1} = \bm{B}\mathbf{u}_t + \mathbf{w}_t,
\end{align}
where $\bm{B}=\bm{I} + \Delta t \bm{M}^{-1}\bm{A}$ and $\bm{w}$ is mean-zero Gaussian with precision $\bm{\Lambda}_{\bm{w}}=(\sigma\Delta t)^{-2}\bm{M}$.
From this we obtain the block-precision (assuming stationarity)
\begin{align}\label{eq:smoothing-precision-fem-euler}
\bm{\Lambda}_{\bm{u}_{(1,\ldots, t)}} = \frac{1}{\sigma^2 \Delta t^2}
\begin{pmatrix}
\bm{M} & -\bm{B}^\top \bm{M} & \bm{0} & \cdots & \bm{0} \\
-\bm{B}^\top \bm{M} & \bm{B}^\top \bm{M} \bm{B} + \bm{M} & -\bm{B}^\top \bm{M} & \cdots & \vdots \\
\bm{0} & -\bm{B}^\top \bm{M} & \bm{B}^\top \bm{M} \bm{B} + \bm{M} & \cdots & \bm{0} \\
\vdots & \vdots & \vdots & \ddots & -\bm{B}^\top \bm{M} \\
\bm{0} & \cdots & \bm{0} &  -\bm{B}^\top \bm{M} & \bm{M}
\end{pmatrix}.
\end{align}
It is evident that we have Markov properties in time.
Spatial Markov properties can also be achieved, attributed to the band-sparsity of \( \bm{A} \) and a (very common) diagonal approximation of \( \bm{M} \):
While \( \bm{M} \) is band-sparse, the non-zero blocks are generally dense, complicating the spatial Markov structure.
By approximating \( \bm{M} \) with its row-sum diagonal, \( \tilde{\bm{M}} = \operatorname{diag}\left(\sum_j M_{i,j}\right) \), we enforce band-sparsity in the resulting blocks, thereby ensuring spatial Markov properties.
This is a common approximation in FEM, which importantly does not increase the error in the solution \citep[Appendix C.5]{lindgren2011explicit}.
In the presence of additional sampling error, the importance of the Markov approximation error in using \( \tilde{\bm{M}} \) is further diminished.

Both the FDM and FEM approaches lead to sparse precision matrices in the Gaussian case, 
implying conditional independence and Markov properties. The key driver for this sparsity is the choice of basis functions that are zero almost everywhere over the discretisation/mesh, combined with the locality of the operators $L$ and $A$.

\subsection{Smoothing: Stochastic heat-equation and Precision Equation \eqref{eq:smoothing-precision-fem-euler}}
\label{app:subsec-stochastic-heat-equation-FEM-spde}

We now showcase explicit usage of Equation \eqref{eq:smoothing-precision-fem-euler} using the 2D stochastic heat equation:
\begin{align}
    \frac{\partial u}{\partial t} - \alpha \laplacian u = \sigma \mathcal{W},
\end{align}
where \( u_t(\bm{x}) \) is the temperature, \( \bm{x} = (x_1, x_2) \) is the spatial position, \( \alpha \) is the thermal diffusivity, and \( \sigma \mathcal{W} \) represents white noise. The spatial domain \( \Omega \) is discretised using a triangular mesh, dividing it into elements \( T_e \) over which basis functions \( \phi_i \) are defined. We approximate \( u(\bm{x}, t) \) by values at the mesh vertices.

The solution requires constructing the matrices $\bm{M}$ and $\bm{A}$ given by Equation \eqref{eq:mass-stiffness-matrices-indices}. Each entry \( M_{ij} \) and \( A_{ij} \) is computed as a sum over the contributions of local elements \( T_e \), which belong to \( \Omega_{ij} \):
\begin{align}
    M_{ij} &= \sum_{e \in \Omega_{ij}} \int_{T_e} \phi_i(\bm{x}) \phi_j(\bm{x}) \, d\bm{x}, \\
    A_{ij} &= \sum_{e \in \Omega_{ij}} \int_{T_e} \alpha \nabla \phi_i(\bm{x}) \cdot \nabla \phi_j(\bm{x}) \, d\bm{x}.
\end{align}

For linear basis functions, the local mass matrix over $T_e$ evaluates to:
\begin{align}
    M^{(e)} = \frac{\text{Area}(T_e)}{12} 
    \begin{pmatrix} 
        2 & 1 & 1 \\ 
        1 & 2 & 1 \\ 
        1 & 1 & 2 
    \end{pmatrix},
\end{align}
with the Markov approximation:
\begin{align}
    \tilde{\bm{M}}^{(e)} = \frac{\text{Area}(T_e)}{3}\bm{I}_3.
\end{align}

The local stiffness matrix represents the Laplacian term and starts as:
\[
A_{ij}^{(e)} = \int_{T_e} \alpha \phi_j(\bm{x}) \laplacian \phi_i(\bm{x}) \, d\bm{x}.
\]
Applying integration by parts to reduce the Laplacian to first-order derivatives, and assuming boundary terms vanish, this becomes:
\[
A_{ij}^{(e)} = \int_{T_e} \alpha \nabla \phi_j(\bm{x}) \cdot \nabla \phi_i(\bm{x}) \, d\bm{x}.
\]

For a triangular element \( T_e \) with vertices \( (x_{1,1}, x_{1,2}) \), \( (x_{2,1}, x_{2,2}) \), and \( (x_{3,1}, x_{3,2}) \), the gradients of the basis functions \( \phi_i \) are constant and given by:
\begin{align}
    \nabla \phi_i = \begin{pmatrix} b_i \\ c_i \end{pmatrix}, \quad 
    b_i = \frac{x_{j,2} - x_{k,2}}{2 \text{Area}(T_e)}, \quad 
    c_i = \frac{x_{k,1} - x_{j,1}}{2 \text{Area}(T_e)},
\end{align}
where \( (i, j, k) \) are cyclic permutations of the triangle vertices. The local stiffness matrix is then:
\begin{align}
    A^{(e)} = \frac{\alpha}{4 \text{Area}(T_e)} 
    \begin{pmatrix} 
        b_1^2 + c_1^2 & b_1 b_2 + c_1 c_2 & b_1 b_3 + c_1 c_3 \\ 
        b_2 b_1 + c_2 c_1 & b_2^2 + c_2^2 & b_2 b_3 + c_2 c_3 \\ 
        b_3 b_1 + c_3 c_1 & b_3 b_2 + c_3 c_2 & b_3^2 + c_3^2 
    \end{pmatrix}.
\end{align}

The global matrices \( \bm{M} \) and \( \bm{A} \) are assembled by summing contributions from all triangular elements \( T_e \) in the domain \( \Omega \). These matrices are sparse, with non-zero entries only for neighbouring vertices in the mesh. The precision of \(\bm{u}_t\) in Equation \eqref{eq:smoothing-precision-fem-euler} retains this sparsity as long as the diagonal approximation \(\tilde{\bm{M}}\) is used.

\section*{Acknowledgments}
I would like to thank Gilson Moura Silva Neto, Max Ramgraber, Tommy Odland, and Tore Kleppe for their valuable feedback on early versions of this paper. Their insights and constructive comments were instrumental in improving the manuscript. Additionally, I am grateful to Equinor for supporting this work.

\bibliographystyle{chicago}
\bibliography{references}

@article{amestoy2004algorithm,
  title={Algorithm 837: AMD, an approximate minimum degree ordering algorithm},
  author={Amestoy, Patrick R and Davis, Timothy A and Duff, Iain S},
  journal={ACM Transactions on Mathematical Software (TOMS)},
  volume={30},
  number={3},
  pages={381--388},
  year={2004},
  publisher={ACM New York, NY, USA}
}

@inproceedings{cuthill1969reducing,
  title={Reducing the bandwidth of sparse symmetric matrices},
  author={Cuthill, Elizabeth and McKee, James},
  booktitle={Proceedings of the 1969 24th national conference},
  pages={157--172},
  year={1969}
}

@article{karypis1997metis,
  title={A fast and high quality multilevel scheme for partitioning irregular graphs},
  author={Karypis, George and Kumar, Vipin},
  journal={SIAM Journal on scientific Computing},
  volume={20},
  number={1},
  pages={359--392},
  year={1998},
  publisher={SIAM}
}

@article{kristensen2015tmb,
  title={{T}{M}{B}: {A}utomatic differentiation and {L}aplace approximation},
  author={Kristensen, Kasper and Nielsen, Anders and Berg, Casper W and Skaug, Hans and Bell, Bradley M},
  journal={Journal of Statistical Software},
  volume={70},
  number={i05},
  year={2016},
  publisher={Foundation for Open Access Statistics}
}

@article{rue2009approximate,
  title={Approximate {B}ayesian inference for latent {G}aussian models by using integrated nested {L}aplace approximations},
  author={Rue, H{\aa}vard and Martino, Sara and Chopin, Nicolas},
  journal={Journal of the {R}oyal {S}tatistical {S}ociety: {S}eries {B} ({S}tatistical {M}ethodology)},
  volume={71},
  number={2},
  pages={319--392},
  year={2009},
  publisher={Wiley Online Library}
}

@article{lunde2022graphspme,
  title={Graph{SPME}: {M}arkov Precision Matrix Estimation and Asymptotic {S}tein-Type Shrinkage},
  author={Lunde, Berent {\AA}nund Str{\o}mnes and Curic, Feda and Sortland, Sondre},
  journal={arXiv preprint arXiv:2205.07584},
  year={2022}
}

@article{lindgren2011explicit,
  title={An explicit link between {G}aussian fields and {G}aussian {M}arkov random fields: the stochastic partial differential equation approach},
  author={Lindgren, Finn and Rue, H{\aa}vard and Lindstr{\"o}m, Johan},
  journal={Journal of the Royal Statistical Society: Series B (Statistical Methodology)},
  volume={73},
  number={4},
  pages={423--498},
  year={2011},
  publisher={Wiley Online Library}
}

@article{lindgren2022spde,
  title={The {SPDE} approach for {G}aussian and non-{G}aussian fields: 10 years and still running},
  author={Lindgren, Finn and Bolin, David and Rue, H{\aa}vard},
  journal={Spatial Statistics},
  pages={100599},
  year={2022},
  publisher={Elsevier}
}

@book{rue2005gaussian,
  title={Gaussian Markov random fields: theory and applications},
  author={Rue, Havard and Held, Leonhard},
  year={2005},
  publisher={Chapman and Hall/CRC}
}

@incollection{oksendal2003stochastic,
  title={Stochastic differential equations},
  author={{\O}ksendal, Bernt},
  booktitle={Stochastic differential equations},
  pages={65--84},
  year={2003},
  publisher={Springer}
}

@book{platen1992numericalspde,
  title={Numerical solution of stochastic differential equations},
  author={Kloeden, Peter E. and Platen, Eckhard},
  year={1992},
  publisher={Springer}
}

@article{rozanov1977markov,
  title={Markov random fields and stochastic partial differential equations},
  author={Rozanov, Ju A},
  journal={Mathematics of the USSR-Sbornik},
  volume={32},
  number={4},
  pages={515},
  year={1977},
  publisher={IOP Publishing}
}

@article{raanes2023review,
  title={Review of ensemble gradients for robust optimisation},
  author={Raanes, Patrick N and Stordal, Andreas S and Lorentzen, Rolf J},
  journal={arXiv preprint arXiv:2304.12136},
  year={2023}
}

@book{hastie2009elements,
  title={The elements of statistical learning: data mining, inference, and prediction},
  author={Hastie, Trevor and Tibshirani, Robert and Friedman, Jerome H},
  volume={2},
  year={2009},
  publisher={Springer}
}

@article{akaike1974new,
  title={A new look at the statistical model identification},
  author={Akaike, Hirotugu},
  journal={IEEE transactions on automatic control},
  volume={19},
  number={6},
  pages={716--723},
  year={1974},
  publisher={Ieee}
}

@article{takeuchi1976distribution,
	title={Distribution of Information Statistics and Validity Criteria of Models},
	author={Takeuchi, Kei},
	journal={Mathematical Science},
	volume={153},
	pages={12--18},
	year={1976}
}

@article{stone1974cross,
	title={Cross-Validatory Choice and Assessment of Statistical Predictions},
	author={Stone, Mervyn},
	journal={Journal of the Royal Statistical Society. Series B (Methodological)},
	pages={111--147},
	year={1974},
	publisher={JSTOR}
}

@article{claeskens2008model,
  title={Model selection and model averaging},
  author={Claeskens, Gerda and Hjort, Nils Lid},
  journal={Cambridge {B}ooks},
  year={2008},
  publisher={Cambridge University Press}
}

@article{konishi1996generalised,
  title={Generalised information criteria in model selection},
  author={Konishi, Sadanori and Kitagawa, Genshiro},
  journal={Biometrika},
  volume={83},
  number={4},
  pages={875--890},
  year={1996},
  publisher={Oxford University Press}
}

@article{evensen1994sequential,
  title={Sequential data assimilation with a nonlinear quasi-geostrophic model using {M}onte {C}arlo methods to forecast error statistics},
  author={Evensen, Geir},
  journal={Journal of Geophysical Research: Oceans},
  volume={99},
  number={C5},
  pages={10143--10162},
  year={1994},
  publisher={Wiley Online Library}
}

@article{burgers1998analysis,
  title={Analysis scheme in the ensemble {K}alman filter},
  author={Burgers, Gerrit and Van Leeuwen, Peter Jan and Evensen, Geir},
  journal={Monthly weather review},
  volume={126},
  number={6},
  pages={1719--1724},
  year={1998},
  publisher={American Meteorological Society}
}

@book{evensen2009data,
  title={Data assimilation: the ensemble Kalman filter},
  author={Evensen, Geir and others},
  volume={2},
  year={2009},
  publisher={Springer}
}

@article{emerick2013ensemble,
  title={Ensemble smoother with multiple data assimilation},
  author={Emerick, Alexandre A and Reynolds, Albert C},
  journal={Computers \& Geosciences},
  volume={55},
  pages={3--15},
  year={2013},
  publisher={Elsevier}
}

@article{houtekamer2005ensemble,
  title={Ensemble {K}alman filtering},
  author={Houtekamer, Peter L and Mitchell, Herschel L},
  journal={Quarterly Journal of the Royal Meteorological Society},
  volume={131},
  number={613},
  pages={3269--3289},
  year={2005},
  publisher={Wiley Online Library}
}

@article{aanonsen2009ensemble,
  title={The ensemble {K}alman filter in reservoir engineering--a review},
  author={Aanonsen, Sigurd I and N{\ae}vdal, Geir and Oliver, Dean S and Reynolds, Albert C and Vall{\`e}s, Brice},
  journal={Spe Journal},
  volume={14},
  number={03},
  pages={393--412},
  year={2009},
  publisher={OnePetro}
}

@article{anderson2003local,
  title={A local least squares framework for ensemble filtering},
  author={Anderson, Jeffrey L},
  journal={Monthly Weather Review},
  volume={131},
  number={4},
  pages={634--642},
  year={2003}
}

@article{evensen2003ensemble,
  title={The ensemble {K}alman filter: {T}heoretical formulation and practical implementation},
  author={Evensen, Geir},
  journal={Ocean dynamics},
  volume={53},
  number={4},
  pages={343--367},
  year={2003},
  publisher={Springer}
}

@article{ott2004local,
  title={A local ensemble {K}alman filter for atmospheric data assimilation},
  author={Ott, Edward and Hunt, Brian R and Szunyogh, Istvan and Zimin, Aleksey V and Kostelich, Eric J and Corazza, Matteo and Kalnay, Eugenia and Patil, DJ and Yorke, James A},
  journal={Tellus A: Dynamic Meteorology and Oceanography},
  volume={56},
  number={5},
  pages={415--428},
  year={2004},
  publisher={Taylor \& Francis}
}

@article{hunt2007efficient,
  title={Efficient data assimilation for spatiotemporal chaos: A local ensemble transform {K}alman filter},
  author={Hunt, Brian R and Kostelich, Eric J and Szunyogh, Istvan},
  journal={Physica D: Nonlinear Phenomena},
  volume={230},
  number={1-2},
  pages={112--126},
  year={2007},
  publisher={Elsevier}
}

@article{hamill2001distance,
  title={Distance-dependent filtering of background error covariance estimates in an ensemble {K}alman filter},
  author={Hamill, Thomas M and Whitaker, Jeffrey S and Snyder, Chris},
  journal={Monthly Weather Review},
  volume={129},
  number={11},
  pages={2776--2790},
  year={2001}
}

@article{houtekamer2001sequential,
  title={A sequential ensemble {K}alman filter for atmospheric data assimilation},
  author={Houtekamer, Peter L and Mitchell, Herschel L},
  journal={Monthly Weather Review},
  volume={129},
  number={1},
  pages={123--137},
  year={2001},
  publisher={American Meteorological Society}
}

@article{kalman1960new,
  title={A new approach to linear filtering and prediction problems},
  author={Kalman, RE},
  journal={Trans. ASME, D},
  volume={82},
  pages={35--44},
  year={1960}
}

@article{kalman1961new,
  title={New results in linear filtering and prediction theory},
  author={Kalman, Rudolph E and Bucy, Richard S},
  journal={Journal of Basic Engineering},
  volume={83},
  number={1},
  pages={95--108},
  year={1961},
  publisher={American Society of Mechanical Engineers}
}

@book{moore1979optimal,
  title={Optimal filtering},
  author={Moore, John Barratt and Anderson, B},
  year={1979},
  publisher={Prentice-Hall New York}
}

@article{rosenblatt1952remarks,
  title={Remarks on a multivariate transformation},
  author={Rosenblatt, Murray},
  journal={The {A}nnals of {M}athematical {S}tatistics},
  volume={23},
  number={3},
  pages={470--472},
  year={1952},
  publisher={JSTOR}
}

@article{ramgraber2023ensemble1,
  title={Ensemble transport smoothing. {P}art I: Unified framework},
  author={Ramgraber, Maximilian and Baptista, Ricardo and McLaughlin, Dennis and Marzouk, Youssef},
  journal={Journal of Computational Physics: X},
  volume={17},
  pages={100134},
  year={2023},
  publisher={Elsevier}
}

@article{ramgraber2023ensemble2,
  title={Ensemble transport smoothing. {P}art II: Nonlinear updates},
  author={Ramgraber, Maximilian and Baptista, Ricardo and McLaughlin, Dennis and Marzouk, Youssef},
  journal={Journal of Computational Physics: X},
  volume={17},
  pages={100133},
  year={2023},
  publisher={Elsevier}
}

@article{baptista2021learning,
  title={Learning non-{G}aussian graphical models via {H}essian scores and triangular transport},
  author={Baptista, Ricardo and Marzouk, Youssef and Morrison, Rebecca E and Zahm, Olivier},
  journal={arXiv preprint arXiv:2101.03093},
  year={2021}
}

@article{marzouk2016introduction,
  title={An introduction to sampling via measure transport},
  author={Marzouk, Youssef and Moselhy, Tarek and Parno, Matthew and Spantini, Alessio},
  journal={arXiv preprint arXiv:1602.05023},
  year={2016}
}

@article{baptista2020representation,
  title={On the representation and learning of monotone triangular transport maps},
  author={Baptista, Ricardo and Marzouk, Youssef and Zahm, Olivier},
  journal={arXiv preprint arXiv:2009.10303},
  year={2020}
}

@article{spantini2019coupling,
  title={Coupling techniques for nonlinear ensemble filtering, arXiv},
  author={Spantini, A and Baptista, R and Marzouk, Y},
  journal={arXiv preprint arXiv:1907.00389},
  volume={30},
  year={2019}
}

@article{spantini2018inference,
  title={Inference via low-dimensional couplings},
  author={Spantini, Alessio and Bigoni, Daniele and Marzouk, Youssef},
  journal={The Journal of Machine Learning Research},
  volume={19},
  number={1},
  pages={2639--2709},
  year={2018},
  publisher={JMLR. org}
}

@article{katzfuss2024scalable,
  title={Scalable Bayesian transport maps for high-dimensional non-Gaussian spatial fields},
  author={Katzfuss, Matthias and Sch{\"a}fer, Florian},
  journal={Journal of the American Statistical Association},
  volume={119},
  number={546},
  pages={1409--1423},
  year={2024},
  publisher={Taylor \& Francis}
}

@article{schafer2021sparse,
  title={Sparse Cholesky Factorization by Kullback--Leibler Minimization},
  author={Sch{\"a}fer, Florian and Katzfuss, Matthias and Owhadi, Houman},
  journal={SIAM Journal on scientific computing},
  volume={43},
  number={3},
  pages={A2019--A2046},
  year={2021},
  publisher={SIAM}
}

@book{van2000asymptotic,
  title={Asymptotic statistics},
  author={Van der Vaart, Aad W},
  volume={3},
  year={2000},
  publisher={Cambridge {U}niversity {P}ress}
}

@inproceedings{huber1967behavior,
  title={The behavior of maximum likelihood estimates under nonstandard conditions},
  author={Huber, Peter J and others},
  booktitle={Proceedings of the fifth Berkeley symposium on mathematical statistics and probability},
  volume={1},
  number={1},
  pages={221--233},
  year={1967},
  organization={Berkeley, CA: University of California Press}
}

@article{tibshirani1996regression,
  title={Regression shrinkage and selection via the lasso},
  author={Tibshirani, Robert},
  journal={Journal of the Royal Statistical Society Series B: Statistical Methodology},
  volume={58},
  number={1},
  pages={267--288},
  year={1996},
  publisher={Oxford University Press}
}

@article{hoerl1970ridge,
  title={Ridge regression: Biased estimation for nonorthogonal problems},
  author={Hoerl, Arthur E and Kennard, Robert W},
  journal={Technometrics},
  volume={12},
  number={1},
  pages={55--67},
  year={1970},
  publisher={Taylor \& Francis}
}

@article{hastie2007forward,
  title={Forward stagewise regression and the monotone lasso},
  author={Hastie, Trevor and Taylor, Jonathan and Tibshirani, Robert and Walther, Guenther},
  journal={Electronic Journal of Statistics},
  volume={1},
  pages={1--29},
  year={2007}
}

@article{friedman2001greedy,
  title={Greedy function approximation: a gradient boosting machine},
  author={Friedman, Jerome H},
  journal={Annals of {S}tatistics},
  pages={1189--1232},
  year={2001},
  publisher={JSTOR}
}

@article{mason1999boosting,
  title={Boosting algorithms as gradient descent},
  author={Mason, Llew and Baxter, Jonathan and Bartlett, Peter and Frean, Marcus},
  journal={Advances in neural information processing systems},
  volume={12},
  year={1999}
}

@inproceedings{lorenz1996predictability,
  author    = {Lorenz, Edward N.},
  title     = {Predictability: A Problem Partly Solved},
  booktitle = {Seminar on Predictability},
  address   = {Shinfield Park, Reading},
  year      = {1995},
  publisher = {ECMWF},
  url       = {https://www.ecmwf.int/node/10829}
}

@article{lorenz1998optimal,
  title={Optimal sites for supplementary weather observations: Simulation with a small model},
  author={Lorenz, Edward N and Emanuel, Kerry A},
  journal={Journal of the Atmospheric Sciences},
  volume={55},
  number={3},
  pages={399--414},
  year={1998},
  publisher={American Meteorological Society}
}

@article{uhlenbeck1930theory,
  title={On the theory of the {B}rownian motion},
  author={Uhlenbeck, George E and Ornstein, Leonard S},
  journal={Physical {R}eview},
  volume={36},
  number={5},
  pages={823},
  year={1930},
  publisher={APS}
}

\end{document}